%% file: alpha2_0.tex
\documentclass[preprint,showpacs,preprintnumbers,amsmath,amssymb]{revtex4}
\usepackage{graphicx}
\usepackage{dcolumn}
\usepackage{bm}

\begin{document}
\preprint{ECT*-05-25}
\title{Non perturbative renormalisation group and momentum dependence of $n$-point functions (II)}

\author{Jean-Paul Blaizot}
  \thanks{CNRS fellow}
  \affiliation{ECT*, Villa Tambosi, strada delle Tabarelle 286, 38050 Villazzano (TN), Italy}
  \author{Ram\'on M\'endez-Galain}
  \email{mendezg@fing.edu.uy}
  \affiliation{Instituto de F\'{\i}sica, Facultad de Ingenier\'{\i}a, J.H.y Reissig 565, 11000
  Montevideo, Uruguay}

\author{Nicol\'as Wschebor}
  \email{nicws@fing.edu.uy}
  \affiliation{
  Instituto de F\'{\i}sica, Facultad de Ingenier\'{\i}a, J.H.y Reissig 565, 11000
  Montevideo, Uruguay}

  \date{\today}
\begin{abstract}

In a companion paper (hep-th/0512317), we have presented an
approximation scheme to solve the Non Perturbative Renormalization
Group equations that allows the calculation of the
 $n$-point functions for arbitrary values of the external momenta.
The method was applied in its leading order to the calculation of
the self-energy of the O($N$) model in the critical regime. The
purpose of the present paper is to extend this study to  the
next-to-leading order of the approximation scheme. This involves the
calculation of the 4-point function at leading order, where new
features arise, related to the occurrence of exceptional
configurations of momenta in the flow equations. These require a
special treatment, inviting us to improve the straightforward
iteration scheme that we originally proposed. The final result for
the self-energy at next-to-leading order exhibits a remarkable
improvement as compared to the leading order calculation. This is
demonstrated by the calculation of the shift $\Delta T_c$, caused by
weak interactions, in the temperature of Bose-Einstein condensation.
This quantity depends on the self-energy at all momentum scales and
can be used as a benchmark of the approximation. The improved
next-to-leading order calculation of the self-energy presented in
this paper leads to excellent agreement with lattice data and  is within 4\% of the exact large $N$ result.\end{abstract}

\maketitle


\def\bfphi{\mbox{\boldmath$\phi$}}
\def\bfvarphi{\mbox{\boldmath$\varphi$}}
\def\bfgamma{\mbox{\boldmath$\gamma$}}
\def\bfalpha{\mbox{\boldmath$\alpha$}}
\def\bftau{\mbox{\boldmath$\tau$}}
\def\bfnabla{\mbox{\boldmath$\nabla$}}
\def\bfsigma{\mbox{\boldmath$\sigma$}}
\def\bfpi{\mbox{\boldmath$\pi$}}

\newcommand \beq{\begin{eqnarray}}
\newcommand \eeq{\end{eqnarray}}

\newcommand \ga{\raisebox{-.5ex}{$\stackrel{>}{\sim}$}}
\newcommand \la{\raisebox{-.5ex}{$\stackrel{<}{\sim}$}}

\def\psib{\psi}
\def\phib{\phi}
\def\r{{\rm r}}
\def\d{{\rm d}}

\def \e {\mbox{e}}

\input epsf


\def\square{\hbox{{$\sqcup$}\llap{$\sqcap$}}}
\def\grad{\nabla}
\def\del{\partial}

\def\frac#1#2{{#1 \over #2}}
\def\smallfrac#1#2{{\scriptstyle {#1 \over #2}}}
\def\half{\ifinner {\scriptstyle {1 \over 2}}
   \else {1 \over 2} \fi}


\def\bra#1{\langle#1\vert}
\def\ket#1{\vert#1\rangle}


\def\simge{\mathrel{%
   \rlap{\raise 0.511ex \hbox{$>$}}{\lower 0.511ex \hbox{$\sim$}}}}
\def\simle{\mathrel{
   \rlap{\raise 0.511ex \hbox{$<$}}{\lower 0.511ex \hbox{$\sim$}}}}


\def\buildchar#1#2#3{{\null\!
   \mathop#1\limits^{#2}_{#3}
   \!\null}}
\def\overcirc#1{\buildchar{#1}{\circ}{}}


\def\slashchar#1{\setbox0=\hbox{$#1$}
   \dimen0=\wd0
   \setbox1=\hbox{/} \dimen1=\wd1
   \ifdim\dimen0>\dimen1
      \rlap{\hbox to \dimen0{\hfil/\hfil}}
      #1
   \else
      \rlap{\hbox to \dimen1{\hfil$#1$\hfil}}
      /
   \fi}


\def\real{\mathop{\rm Re}\nolimits}     
\def\imag{\mathop{\rm Im}\nolimits}     

\def\tr{\mathop{\rm tr}\nolimits}       
\def\Tr{\mathop{\rm Tr}\nolimits}       
\def\Det{\mathop{\rm Det}\nolimits}     

\def\mod{\mathop{\rm mod}\nolimits}     
\def\wrt{\mathop{\rm wrt}\nolimits}     


\def\TeV{{\rm TeV}}                     
\def\GeV{{\rm GeV}}                     
\def\MeV{{\rm MeV}}                     
\def\KeV{{\rm KeV}}                     
\def\eV{{\rm eV}}                       

\def\mb{{\rm mb}}                       
\def\mub{\hbox{$\mu$b}}                 
\def\nb{{\rm nb}}                       
\def\pb{{\rm pb}}                       

%
%

\def\picture #1 by #2 (#3){
  \vbox to #2{
    \hrule width #1 height 0pt depth 0pt
    \vfill
    \special{picture #3} 
    }
  }

\def\scaledpicture #1 by #2 (#3 scaled #4){{
  \dimen0=#1 \dimen1=#2
  \divide\dimen0 by 1000 \multiply\dimen0 by #4
  \divide\dimen1 by 1000 \multiply\dimen1 by #4
  \picture \dimen0 by \dimen1 (#3 scaled #4)}
  }

\def\centerpicture #1 by #2 (#3 scaled #4){
   \dimen0=#1 \dimen1=#2
    \divide\dimen0 by 1000 \multiply\dimen0 by #4
    \divide\dimen1 by 1000 \multiply\dimen1 by #4
         \noindent
         \vbox{
            \hspace*{\fill}
            \picture \dimen0 by \dimen1 (#3 scaled #4)
            \hspace*{\fill}
            \vfill}}


\def\figfermass{\centerpicture 122.4mm by 32.46mm
 (fermass scaled 750)}

%

\section{Introduction}
The development of  non-perturbative methods is essential to be able
to deal with a large variety of  problems in which the absence of a
small parameter prevents one to build solutions in terms of a
systematic expansion. Among such methods, the non perturbative
renormalization group (NPRG)
\cite{Wetterich93,Ellwanger93,Tetradis94,Morris94,Morris94c} stands
out as a very promising tool, suggesting new approximation schemes
which are not easily formulated in other, more conventional,
approaches in field theory or many body physics. The NPRG has been
applied successfully to a variety of physical problems, in condensed
matter, particle or nuclear physics (for  reviews, see e.g.
\cite{Bagnuls:2000ae,Berges02,Canet04}). In most of these problems
however, the focus is on long wavelength modes  and  the solution of
the NPRG equations involves generally a derivative expansion which
only allows for the determination of the $n$-point functions and
their derivatives at small external momenta (vanishing momenta in
the case of critical phenomena). In many situations, this is not
enough: a full knowledge of the momentum dependence of the
correlation functions is needed to calculate the quantities of
physical interest.

For this reason, in ref.~\cite{alpha1}, we have presented an
approximation scheme to solve the NPRG  equations that allows the
calculation of the
 $n$-point functions for arbitrary values of the external momenta.
The method was applied in its leading order to the calculation of
the self-energy of the O($N$) model in the critical regime. The
purpose of the present paper is to extend this study to  the
next-to-leading order of the approximation scheme. This involves the
calculation of the 4-point function at leading order, where new
features arise. In particular  we need to deal with exceptional
configurations of momenta that enter the flow equations. Because of
these, the straightforward iteration scheme   proposed in
ref.~\cite{alpha1} yields some unphysical features in  the 4-point
function. After having identified the origin of the problem, we
shall show how it can be cured by a proper treatment of the flow
equation in the channel where the exceptional momenta matter. The
final result for  the self-energy at next-to-leading order exhibits
a significant improvement as compared to the leading order
calculation. In particular the calculation of the the shift $\Delta
T_c$ in the transition temperature of the weakly repulsive Bose gas
\cite{club}, a quantity which is very sensitive to all momentum
scales and which is used as a benchmark of the approximation, is now
in excellent agreement with the available lattice data, and within
4\% of the exact large $N$ result. Note that preliminary results
concerning the calculation of $\Delta T_c$ at next-to-leading order
have been presented in ref.~\cite{Blaizot:2004qa}. These results
were obtained without the improvements just alluded to.

This paper is a sequel of ref.~\cite{alpha1}, and should be read in
conjunction with it (hereafter ref.~\cite{alpha1} will be referred
to as paper I, and the prefix I in equation labels will refer  to
equations in paper I). As in paper I we shall focus  the discussion
on the O($N$) model, although most of the arguments  have a wider
range of applicability. Thus, we shall consider  a scalar
$\varphi^4$ theory in $d$ dimension with $O(N)$ symmetry: \beq
\label{classicalaction} {\cal S} = \int {\rm d}^{d}x\, \left\lbrace{
1 \over 2} \left[ \nabla \varphi (x) \right]^2+{1 \over 2}r
\varphi^2 (x)+{u \over 4!} \left[ \varphi^2(x) \right]^2
\right\rbrace , \eeq The field $\varphi(x)$ has $N$ real components
$\varphi_i(x)$, with $i=1,\cdots,N$.

The basic equation of the NPRG is the flow equation for the
effective action $\Gamma_\kappa[\phi]$  which interpolates between
the classical action ${\cal S}$ and the full effective action
$\Gamma[\phi]$ ($\phi$ is the expectation value of the field) as the
parameter $\kappa$ varies from the microscopic scale $\Lambda$ down
to zero. The flow equation for the effective action
$\Gamma_\kappa[\phi]$ reads
\cite{Wetterich93,Ellwanger93,Tetradis94,Morris94,Morris94c}:
\beq\label{mastereq}
\partial_\kappa \Gamma_\kappa[\phi]=\frac{1}{2}\,{\rm tr}\int \frac{d^dq}{(2\pi)^d}
\,\partial_\kappa R_\kappa(q^2)
\left[\Gamma_\kappa^{(2)}+R_\kappa\right]^{-1}_{q,-q}, \eeq where
the trace runs over the O($N$) indices, $\Gamma_\kappa^{(2)}$ is the
second derivative of $\Gamma_\kappa$ with respect to $\phi$, and
$R_\kappa(q)$ is the regulator chosen, as in paper I, of the form
\cite{Litim} \beq\label{reg-litim} R_\kappa(q)\propto (\kappa^2-q^2)
\theta(\kappa^2-q^2). \eeq The role of the regulator is to suppress
the fluctuations with momenta $q\simle \kappa$, while leaving
unaffected those with $q\simge \kappa$.

The flow equations for the various $n$-point functions are obtained by taking functional derivatives with respect to $\phi$  of the equation for $\Gamma_\kappa[\phi]$  (see eq.~(I.8)). In particular, the  self-energy $\Sigma(\kappa;p)$ is obtained by integrating the flow equation (I.9):
\begin{equation}\label{eq:dGamma2}
 \hspace{-1cm}\partial_\kappa
\Gamma^{(2)}_{12}(\kappa;p)\equiv \delta_{12}\del_\kappa \Sigma(\kappa; p)= -\frac{1}{2}\int
\frac{d^dq}{(2\pi)^d} \,\partial_\kappa R_\kappa(q)\,G^2(\kappa;q)
\,\Gamma^{(4)}_{12ll}(\kappa;p,-p,q,-q),
\end{equation}
with the inverse  propagator  given by \beq
\label{Ginverse}G^{-1}(\kappa,q)=q^2+\Sigma(\kappa;q)+R_\kappa(q).
\eeq In eq.~(\ref{eq:dGamma2}), and later in this paper, we often
denote the O($N$) indices simply by numbers $1,2,$ etc., instead of
$i_1, i_2,$ etc.,  in order to alleviate the notation.

The flow equations for the $n$-point functions constitute an
infinite hierarchy of coupled equations (for example, eq.
(\ref{eq:dGamma2}) for the 2-point function contains in its r.h.s.
the 4-point function). In paper I we have proposed a strategy to
solve this hierarchy  by following an iterative procedure. This
starts with an \emph{initial ansatz} for the $n$-point functions to
be used in the right hand side of the flow equations.  Integrating
the flow equation of a given $n$-point function gives then the {\it
leading order} (LO) estimate for that $n$-point function. Inserting
the leading order of the $n$-point functions thus obtained in the
right hand side of the flow equations and integrating gives then the
{\it next-to-leading order} (NLO) estimate of the $n$-point
functions. And so on.

Recall that there is no small parameter controlling the convergence
of the process, and the terminology LO, NLO, refers merely to the
number of iterations involved in the calculation of the $n$-point
function considered. Since the calculations become increasingly
complicated as the number of iterations increases, the success of
the procedure relies crucially on the quality of the initial ansatz.
A major task then is to
 construct such a good initial ansatz.

The equations are solved starting at the bottom of the hierarchy,
that is, with the equation for the 2-point function which involves,
in its right hand side, the 2-point function (through the
propagator), and the 4-point function. As initial ansatz for the
propagator, we take the propagator of a modified version of the
derivative expansion that we called the LPA'; it is given by (see
paper I, sect. II): \beq\label{propLPA'}
G_{LPA'}^{-1}(\kappa;q)=Z_\kappa q^2+m_\kappa^2+R_\kappa(q), \eeq
where the field renormalization factor $Z_\kappa$ and the running
mass $m_\kappa$ are obtained by solving the LPA' equations (see
paper I, sect. IIB). The initial ansatz for the 4-point function is
given explicitly in paper I, sect.~III, and is obtained as the
solution of an approximate equation (see eq.~(I.68) and
eq.~(\ref{gamma40new}) below). For more clarity, we shall
distinguish in this paper the initial ansatz for $ \Gamma^{(4)}$ by
a tilde. For consistency, we shall use a similar notation for the
initial ansatz for the propagator, i.e., we shall set $\tilde
G\equiv G_{LPA'}$. Summarizing, the leading order self-energy is
given by: \beq\label{SigmaLO}
 \delta_{12}\del_\kappa \Sigma_{LO}(\kappa; p)= -\frac{1}{2}\int
\frac{d^dq}{(2\pi)^d} \,\partial_\kappa R_\kappa(q)\,\tilde
G^2(\kappa;q) \,\tilde\Gamma^{(4)}_{12ll}(\kappa;p,-p,q,-q). \eeq
 Note that, as compared to eq.~(\ref{eq:dGamma2}), eq.~(\ref{SigmaLO}) is now a trivial  flow equation
since all quantities in the r.h.s. are known quantities: $\Sigma_{LO}(\kappa; p)$ is simply obtained by integrating the r.h.s. with respect to $\kappa$. The leading
order self-energy has been studied  in detail in paper I.

As stated earlier, the purpose of the present paper is to calculate
$\Sigma_{NLO}(p)$, the self-energy at next-to-leading order. To do so,
we need to use in the r.h.s of eq.~(\ref{eq:dGamma2}) the leading
order expressions for both the propagator and the 4-point function.
The leading order propagator is obtained from eq.~(\ref{Ginverse})
with the self-energy  $\Sigma_{LO}$. Getting the leading order
expression for the 4-point function
$\Gamma^{(4)}_{12ll}(\kappa;p,-p,q,-q)$ will be the main task of
this paper; it is presented in sect.~\ref{gamma4LO}. First, in
sect.~\ref{gamma4LOd}, we follow the procedure outlined above, i.e.,
replace in the r.h.s. of the flow equation for the 4-point function,
eq.~(\ref{ecgamma4LO})
 below, the initial ansatz for the propagator, the 4- and the 6-point
functions. The initial ansatz for the 6-point function is obtained
by following the same strategy as that used in paper I, sec.~III, in
order to construct the initial ansatz for the 4-point function.
Although conceptually straightforward, this is technically more
involved and the details are presented in app.~\ref{sec:gamma6}.
Then, in sec.~\ref{gamma4LOi}, we  present an improved procedure to
calculate $\Gamma^{(4)}_{12ll}(\kappa;p,-p,q,-q)$ at LO, which copes
properly with the difficulties related to the exceptional
configuration of momenta.  The properties of $\Sigma_{NLO}$  are
presented in sect.~\ref{sec-selfNLO}, together with the result of
the calculation of $\Delta\langle \varphi^2\rangle$, which we use, as
we have recalled earlier, as a benchmark of the approximation
scheme. The last section summarizes the results, and points to
further improvements of the approximation scheme that we have
already started to implement \cite{PLB,BMWn}.

\input{alpha2_1}

\acknowledgements
Authors R. M-G and N. W are grateful for the hospitality of  the ECT* in Trento where part of this work was carried out. 
\bibliographystyle{unsrt}

\end{document}

%% file: alpha2_1.tex
\section{The 4--point function at LO \label{gamma4LO}}

\begin{figure}[t!]
\begin{center}
\includegraphics*[scale=0.8]{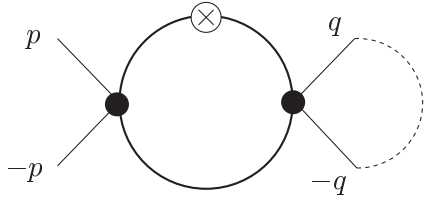}
\end{center}
\caption{\label{ti} Diagrammatic illustration of the contribution
of  the $s$-channel to the flow of the 4-point function (the second line in eq.~(\ref{ecgamma4LO})). The dotted line indicates the loop integral involving
this contribution of $\Gamma^{(4)}$ in the calculation of the
self-energy in NLO, i.e., $\Sigma_{NLO}^{[s]}(p)$ (see sect.~III).}
\end{figure}

\begin{figure}[t!]
\begin{center}
\includegraphics*[scale=0.8]{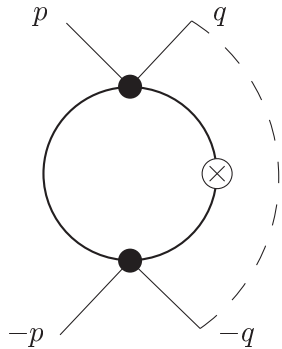}
\end{center}
\caption{\label{sunset} Diagrammatic illustration of the
contribution of the $t$ and $u$ channels to the flow of
the 4-point function (third and fourth  lines
in eq.~(\ref{ecgamma4LO})). The dashed  line
indicates the loop integral involving this contribution of
$\Gamma^{(4)}$ in the calculation of  the self-energy in NLO, i.e.,
 $\Sigma_{NLO}^{[t+u]} (p)$ (see sect.~III).}
\end{figure}

\begin{figure}[t!]
\begin{center}
\includegraphics*[scale=0.8]{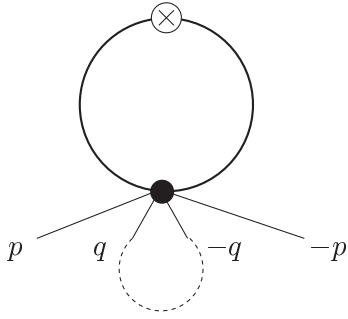}
\end{center}
\caption{\label{octupus} Diagrammatic illustration of the
contribution of the 6-point function to the flow of
the 4-point function (last line in
eq.~(\ref{ecgamma4LO})). The dotted line indicates the
loop integral involving this contribution of $\Gamma^{(4)}$ in the
calculation of the self-energy in NLO, i.e., $\Sigma_{NLO}^{[6]}(p)$ (see sect.~III).}
\end{figure}

The flow equation for  the  the 4-point function in vanishing field
reads (see e.g. eq.~(I.11)):
\begin{eqnarray}
\label{ecgamma4LO}
&&\partial_\kappa\Gamma^{(4)}_{12ll}(\kappa;p,-p,q,-q)=
\int \frac{d^dq'}{(2\pi)^d}\partial_\kappa R_\kappa(q')  G^2(\kappa;q') \nonumber \\
&&\hspace{.5cm}\times\left\lbrace
G(\kappa;q') \Gamma^{(4)}_{12ij}(\kappa;p,-p,q',-q')
 \Gamma^{(4)}_{llij}(\kappa;q,-q,-q',q') \right.\nonumber \\
&&\hspace{.5cm}+
G(\kappa;q'+p+q) \Gamma^{(4)}_{1lij}(\kappa;p,q,q',-q'-p-q)
 \Gamma^{(4)}_{2lij}(\kappa;-p,-q,-q',q'+p+q) \nonumber \\
&&\hspace{.5cm}\left.+
G(\kappa;q'+p-q) \Gamma^{(4)}_{1lij}(\kappa;p,-q,q',-q'-p+q)
 \Gamma^{(4)}_{l2ij}(\kappa;q,-p,-q',q'-q+p) \right\rbrace \nonumber \\
&&\hspace{.5cm}-\frac{1}{2}\int \frac{d^dq}{(2\pi)^d}\partial_\kappa
R_\kappa(q')  G^2(\kappa;q')
 \Gamma^{(6)}_{12llmm}(\kappa;p,-p,q,-q,q',-q').
\end{eqnarray}
We have specified here the O($N$) indices and the momenta of the particular 4-point function $\Gamma_{12ll}^{(4)}(\kappa;p,-p,q,-q)$   that is
needed for the calculation of  $\Sigma_{NLO}(p)$ (see e.g. eq.~(\ref{eq:dGamma2})).  Following the
terminology introduced in paper I, we refer to the second line of
eq.~(\ref{ecgamma4LO}) as to the $s$-channel contribution, while
the third and fourth lines are respectively the  $t$ and $u$-channel
contributions. Note that the contribution of the $u$-channel differs
from that of the $t$-channel solely by the change of sign of $q$. A
graphical illustration of these various contributions is given in
figs.~\ref{ti}-\ref{octupus}.

As stated above, the leading order expression for $\Gamma^{(4)}$ is
obtained by substituting in the r.h.s. of eq.~(\ref{ecgamma4LO}) the
initial ansatz  for $G$, $\Gamma^{(4)}$ and $\Gamma^{(6)}$.  In
fact, we shall proceed with a further simplification which consists
in setting $q'=0$ in the vertices of the r.h.s. of
eq.~(\ref{ecgamma4LO}). For $\tilde\Gamma^{(4)}$, this  is justified
by the fact that the initial anstaz $\tilde\Gamma^{(4)}$ varies
little when the momenta on which it depends are varied by an amount
smaller than $\kappa$.  This property has been explicitly assumed in
the construction of $\tilde\Gamma^{(4)}$ in paper I. It has also
been tested quantitatively in the calculation of the leading order
self-energy (see fig.~15  in paper I  and the discussion at the end
of sect.~IV of paper I).   We assume that this property also
holds for the initial ansatz
$\tilde\Gamma^{(6)}_{12llmm}(\kappa;p,-p,q,-q,q',-q')$.  In fact,
for this latter function we shall also set $q=0$, which can be
justified as follows. Note first that  $\Gamma^{(4)}(p,-p,q,-q)$
will eventually be used in the calculation of $\Sigma_{NLO}(p)$, and
in this calculation $q\simle \kappa$. As we explained in paper I,
sect.~IIIA, setting $q=0$  is then well justified when $p\simle
\kappa$, because in that case all the momenta are smaller than
$\kappa$, and $\Gamma^{(6)}$ is well approximated by the LPA'; it is
also justified  when $p \gg \kappa$ since then $q$ is negligible
compared to $p$. It is only in the small integration region  $q\sim
\kappa$ that the approximation could be less accurate.  Observe finally
that the contribution of $\Gamma^{(6)} $ is negligible unless $ p\ll
\kappa_c$. Thus, in line with approximation ${\cal A}_1$ of paper
I, we shall,  in the r.h.s. of eq.~(\ref{ecgamma4LO}), set $q'=0$ in
the 4-point functions $\tilde\Gamma^{(4)}$ and $q=q'=0$ in
$\tilde\Gamma^{(6)}$. We then arrive at the following  simplified
equation:
\begin{eqnarray}
\label{4pointLOapp}
&&\kappa\partial_\kappa \Gamma^{(4)}_{12ll}(\kappa;p,-p,q,-q)=
I_d^{(3)}(\kappa)\tilde\Gamma^{(4)}_{12ij}(\kappa;p,-p,0,0)\tilde\Gamma^{(4)}_{llij}(\kappa;q,-q,0,0) \nonumber \\
&&\hspace{.5cm}+J_d^{(3)}(\kappa;p+q)\tilde\Gamma^{(4)}_{1lij}(\kappa;p,q,0,-p-q)
\tilde\Gamma^{(4)}_{2lij}(\kappa;-p,-q,0,p+q)
\nonumber \\
&&\hspace{.5cm}+J_d^{(3)}(\kappa;p-q)\tilde\Gamma^{(4)}_{1lij}(\kappa;p,-q,0,-p+q)
\tilde\Gamma^{(4)}_{2lij}(\kappa;-p,q,0,p-q) \nonumber \\
&&\hspace{.5cm}-\frac{1}{2}I_\d^{(2)}(\kappa)\tilde\Gamma^{(6)}_{12llmm}(\kappa;p,-p,0,0,0,0),
\end{eqnarray}
where  the functions $I_d^{(n)}(\kappa)$ and $J_d^{(n)}(\kappa;p)$
are defined  in eqs.~(I.42) and (I.55), respectively.
The  construction of the initial ansatz
$\tilde\Gamma^{(6)}_{12llmm}(\kappa;p,-p,0,0,0,0)$ requires the
solution of an approximate
 flow equation which is obtained by following the same three approximations
that are used in paper I, sect. III,  to get $\tilde\Gamma^{(4)}$.
This is presented in app.~\ref{sec:gamma6}. The explicit traces of
products of the  functions $\tilde\Gamma^{(4)}$ appearing in
eq.~(\ref{4pointLOapp}) are given in app.~B.

Eq.~(\ref{4pointLOapp}) will be used to calculate $\Gamma^{(4)}$ at
LO. Note that, as was the case for eq.~(\ref{SigmaLO}) giving
$\Sigma_{LO}$, eq.~(\ref{4pointLOapp}) is now a trivial flow
equation: all quantities in its r.h.s. are known quantities.  We
shall refer to this calculation of $\Gamma^{(4)}$ at LO, which
follows strictly the scheme proposed in paper I, as to the ``direct
procedure''. The results obtained in this way are discussed in the
next subsection. We shall see there that this procedure yields
unphysical features in some specific situations, whose origin will
be discussed. An improvement on the direct procedure will then be
proposed in the following subsection.

Before proceeding further, let us mention that we have used the LO
estimate of $\Gamma_{12ll}^{(4)}(\kappa,p,-p,q,-q)$ obtained in the
direct procedure to perform a consistency check of approximation
${\cal A}_1$ used in order to obtain eq.~(\ref{4pointLOapp}). We
have verified that $\Gamma_{12ll}^{(4)}(\kappa,p,-p,q,-q)$ varies
little as $q$ varies in the range $q < \kappa $, which is the range
relevant for  the calculation of $\Sigma_{NLO}(p)$. Only in a small
region where $\kappa$ is of order  $ p \ll \kappa_c$, can the function change by
as much as 10\% when $q$ goes from 0 to $\kappa$. In all the other
regions the variation is less than 1\%. In the calculation of
$\Delta\langle\varphi^2\rangle$ that will be reported in the next
section, one needs $\Sigma_{NLO}(p)$ for values of $p$ around
$\kappa_c$: there, the approximation ${\cal A}_1$ is indeed
excellent.  However, at very small momenta, the error due to
this approximation on the magnitude of $\Sigma_{NLO}(p)$ can be
large.  But in this region the approximation ${\cal A}_1$  is not
the dominant source of error anyway: in particular, any error on the
exponent $\eta$ will translate into a large relative error on the
magnitude of $\Sigma_{NLO}(p)$.

\subsection{Direct procedure \label{gamma4LOd}}

As we have just discussed, since all quantities in the r.h.s. of
eq.~(\ref{4pointLOapp}) are known,
$\Gamma_{12ll}^{(4)}(\kappa;p,-p,q,-q)$ at LO is obtained by simply
integrating  the r.h.s. of eq.~(\ref{4pointLOapp}) between $\Lambda$
and $\kappa$,  and adding the bare value of $\Gamma^{(4)}$ (i.e.,
the value of $\Gamma^{(4)}$ at the microscopic scale $\Lambda$):
\beq\label{init-cond}
\Gamma_{12ll}^{(4)}(\kappa=\Lambda;p,-p,q,-q)=( N+2) \; g_{\Lambda}
\; \delta_{12} \eeq with $g_{\Lambda}=  u/3$,  $u$ being the
parameter of the classical action (\ref{classicalaction}) (see paper
I, sect. IIB).

\begin{figure}[t!]
\begin{center}
\includegraphics*[width=8 cm,angle=-90]{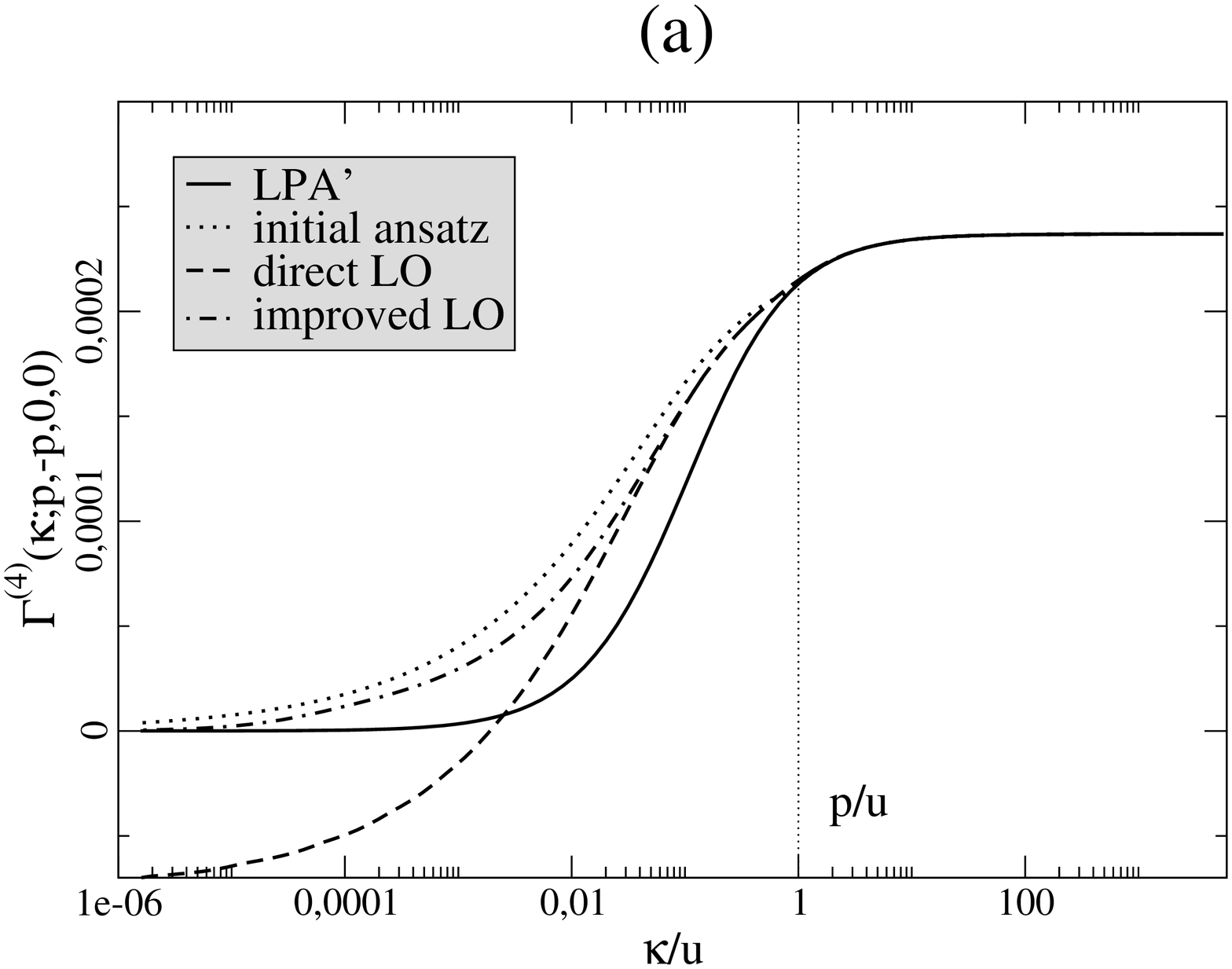}
\hspace{1cm}
\includegraphics*[width=8cm,angle=-90]{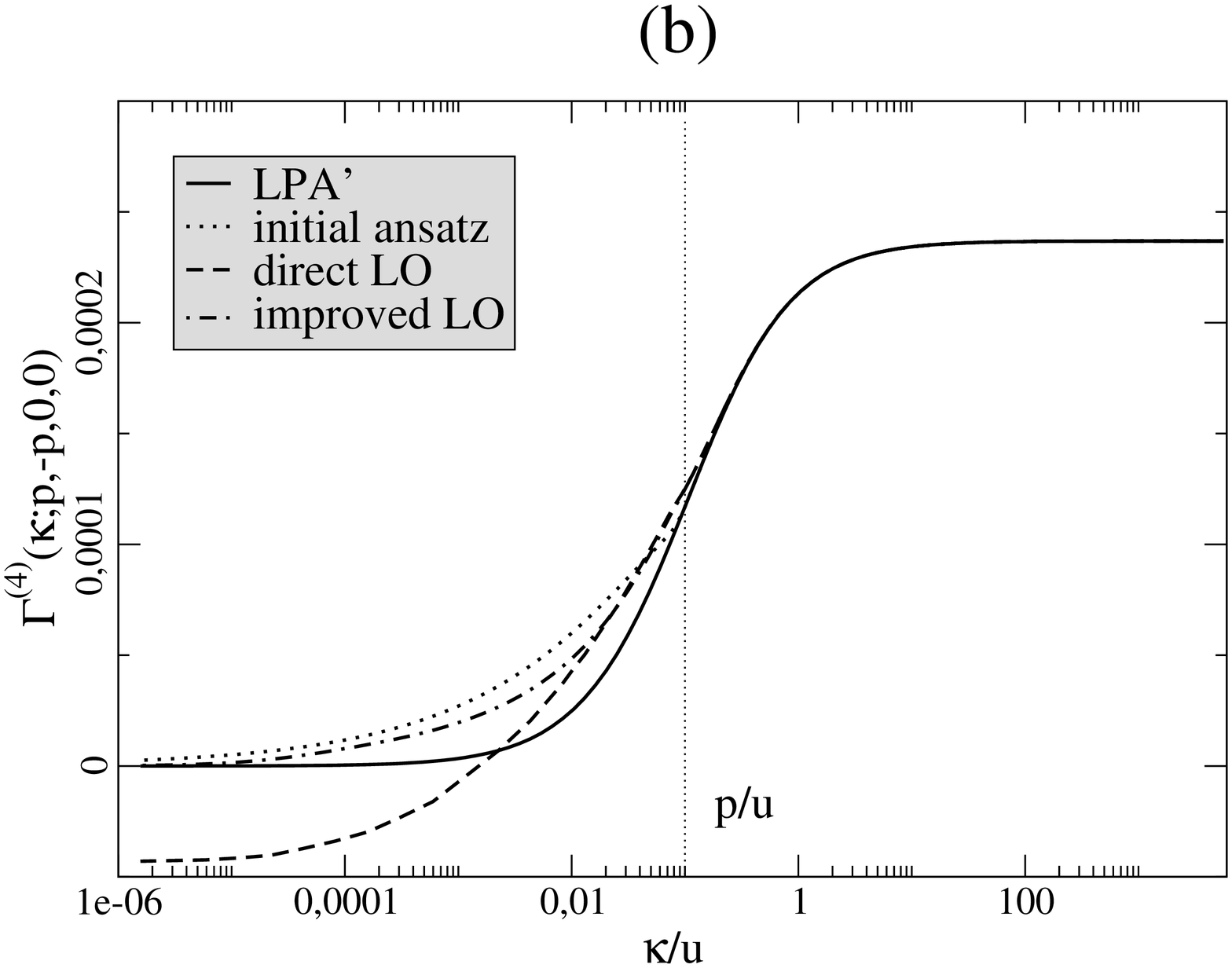}
\end{center}
\caption{\label{gamma4_p=u} The 4-point function $\Gamma^{(4)}
(\kappa;p,-p,0,0)$ (in units of $\Lambda$) as a function of
$\kappa/u$: the four curves represent respectively the LPA' (full
line), the initial ansatz (dotted line), the direct LO (dashed line)
and the improved LO (dot-dashed line). The calculation is done for
(a): $p= \; u$, and (b): $p=\;u/10$.}
\end{figure}

The LO value of
$\Gamma^{(4)}_{12ll}(\kappa,p,-p,0,0)$  thus obtained is compared  with both the initial ansatz and the LPA' result,
$(N+2)g_\kappa$,  in fig.~\ref{gamma4_p=u}
 below. When $\kappa \simge p$ one
expects the LPA' to be a good approximation, and indeed the three
curves almost coincide (the initial ansatz  is by construction identical to
the LPA' for large $\kappa$). When $\kappa$ goes to values smaller
than $p$, the flow of the 4-point function
 is
expected to be slower than that of the LPA' since, generally,
momenta in the propagators tend to suppress the flow.
 Fig.~\ref{gamma4_p=u}  shows that it is indeed the case (in  the initial ansatz this feature is implemented in a sharp
manner with the help of $\Theta$-functions depending on a parameter
$\alpha$ (see paper I, sect. III, and eq.~(\ref{gamma40new})
below)).  As shown by fig.~\ref{gamma4_p=u}, as $\kappa/p$ decreases
further, the LO curve remains close to that corresponding to the
initial ansatz but when $\kappa/p$ becomes too small, eventually the
two curves deviate: while the initial ansatz goes to zero as
$\kappa\to 0$, the LO curve goes to a negative value. As we shall
explain shortly this is an unphysical feature of the LO in the
direct procedure.

At this point, it is instructive to recall the form of the
approximate flow equation whose solution is the initial ansatz
$\tilde\Gamma^{(4)}$.  This equation is established  in paper I,
sect.~III, and reads (see eq.~(I.68)):
\begin{eqnarray}
\label{gamma40new}\small
&&\hspace{-0.6cm}\kappa\partial_\kappa\tilde\Gamma^{(4)}_{12ll}(\kappa;p,-p_,q,-q)=I_d^{(3)}(\kappa)\,
(1-F_\kappa)
  \nonumber \\
&&\hspace{.5cm}\times\left\lbrace
\tilde\Gamma^{(4)}_{12ij}(p,-p,0,0)\tilde\Gamma^{(4)}_{llij}
(q,-q,0,0) \right.\nonumber \\
&&\hspace{.5cm}+\;\;\;\Theta\left({\kappa^2}-\alpha^2
{(p+q)^2}\right) \tilde\Gamma^{(4)}_{1lij}(p,q,0,-p-q)
\tilde\Gamma^{(4)}_{2lij}(-p,-q,0,p+q) \nonumber \\
&&\hspace{.5cm}\left.+\;\;\;\Theta\left({\kappa^2}-\alpha^2
{(p-q)^2} \right)\tilde\Gamma^{(4)}_{1lij}(p,-q,0,-p+q)
\tilde\Gamma^{(4)}_{2lij}(-p,q,0,p-q) \right\rbrace .\nonumber\\
\end{eqnarray}\normalsize
The function $F_\kappa$ is defined in eq.~(I.44). It measures
(approximately) the magnitude of the contribution of the 6-point
function  relative to that of the terms containing the 4-point
functions. Note the similarity between the r.h.s. of
eq.~(\ref{4pointLOapp}) giving the LO expression of $\Gamma^{(4)}$
and the r.h.s. of eq.~(\ref{gamma40new}) for the initial ansatz
$\tilde\Gamma^{(4)}$: the main differences are the replacement of
the $\Theta$-functions of eq.~(\ref{gamma40new}) by the proper loop
integrals in eq.~(\ref{4pointLOapp}), and a more accurate treatment
of the contribution of the 6-point function in
eq.~(\ref{4pointLOapp}). This observation leads us to expect that
$\Gamma^{(4)}$ at LO should not differ much from
$\tilde\Gamma^{(4)}$, a necessary condition for the validity of the
iteration procedure. As shown by fig.~\ref{gamma4_p=u}, this
expectation is fulfilled, except for small values of $\kappa$: while
the LO $\Gamma^{(4)}$ goes to a negative value when $\kappa\to 0$,
the initial ansatz $\tilde\Gamma^{(4)}(\kappa;p,-p,0,0)$ vanishes
(one reads from eq.~(I.99) that in the limit $\kappa \to 0$,
$\tilde\Gamma^{(4)}(\kappa;p,-p,0,0)$ goes to zero like
$g_\kappa^{(N+2)/(N+8)}$).  There is in fact a major difference
between eq.~(\ref{gamma40new}) for the initial ansatz and
eq.~(\ref{4pointLOapp}) giving $\Gamma^{(4)}$ at leading order:
while in eq.~(\ref{gamma40new})  the unknown function
$\tilde\Gamma^{(4)}(\kappa;p,-p,0,0)$ sits in the r.h.s, this is not
so in eq.~(\ref{4pointLOapp}), as we have already emphasized. Thus
the structures of eqs.~(\ref{gamma40new}) and (\ref{4pointLOapp})
are different, and only eq.~(\ref{gamma40new}) captures the
essential feature that guarantees the vanishing of
$\tilde\Gamma^{(4)}(\kappa;p,-p,0,0)$  as $\kappa\to 0$, which, as
we shall show now, is a property of the exact solution of the flow
equation. 

To do so, we shall consider  eq.~(\ref{ecgamma4LO}) for
$\Gamma^{(4)}(\kappa;p,-p,0,0)$ and study the behaviour of its
solution in the limit when $\kappa \to 0$. Observe
first that the $s$-channel controls the flow when $\kappa\simle p$:  indeed, in this channel, the exceptional configuration
of momenta $(p,-p)$ makes the loop integral in the r.h.s. of the
flow equation independent of $p$. This is manifest in
eq.~(\ref{4pointLOapp}): the loop integral reduces to the momentum
independent function $I_d^{(3)}(\kappa)$, while in the other
channels the $p$-dependent functions $J_d^{(3)} (p\pm q)$ enter
(this is also obvious in eq.~(\ref{gamma40new}) where the functions
$J_d^{(3)} (p\pm q)$ are approximated using $\Theta$-functions). It
follows that in the $s$-channel, the external momentum $p$ does not
contribute to stabilize the flow whenever $\kappa\simle p$, as it
does in the other channels: the flow continues all the way down to
$\kappa=0$. When treated correctly, this is what induces the
vanishing of $\Gamma^{(4)}(\kappa;p,-p,0,0)$. To show this latter
point, we focus on the contribution of the $s$-channel and, in the
r.h.s. of eq.~(\ref{4pointLOapp}) we keep
$\Gamma^{(4)}(\kappa;p,-p,0,0)$ as an unknown function (instead of
replacing it by the initial ansatz
$\tilde\Gamma^{(4)}(\kappa;p,-p,0,0))$. We also temporarily neglect
the contribution of the other two channels, and also the
contribution of the  6-point function. Then, eq.~(\ref{ecgamma4LO})
takes the following simple form (here we focus on the structure of
the equation, dropping the O($N$) indices for simplicity, as well as
explicit reference to the external momenta; a more precise equation
is written in the next subsection): \beq\label{eqnapproxforgamma}
\kappa\partial_\kappa \Gamma(\kappa) = g_\kappa I_3^{(3)} (\kappa)
\, \Gamma(\kappa),\eeq where we have replaced
$\tilde\Gamma^{(4)}(\kappa;0,0,0,0)$ by the LPA vertex $g_\kappa$
(see eq.~(\ref{GammaLPA}) below). In the limit $\kappa\to 0$ (see
paper I), $g_\kappa\sim Z_\kappa^2\; \kappa^{4-d}$
 and $I_d^{(3)} (\kappa)\sim \kappa^{d-4}/Z_\kappa^2$, so that
 $g_\kappa I_d^{(3)} (\kappa)\to \xi$, where $\xi$ is a
 positive
 constant. It then follows form eq.~(\ref{eqnapproxforgamma}) that $\Gamma(\kappa)
 \sim
 \kappa^\xi$  as $\kappa\to 0$, and hence vanishes as $\kappa$
 vanishes. Turning now to the contributions of the  $t$ and $u$-channels, we note that
 these are more regular than the $s$-channel contribution (this can be seen for instance
from the fact that the ratio
$J_d^{(3)}(\kappa;p)/I_d^{(3)}(\kappa)\to 0$ as $\kappa/p \to 0$;
see paper I, fig.~9). Similarly,
 the contribution of the 6-point function is proportional
to $I_d^{(2)}$ and $I_d^{(2)}/I_d^{(3)}\sim \kappa^{2-2\eta}$ as
$\kappa \to 0$.  Using these properties, one can write  the
equation for the 4-point function at small $\kappa$
 in the schematic form  \beq\label{fulleqnforGamma}
\kappa\del_\kappa\Gamma(\kappa)=\xi\,\Gamma(\kappa)+\Phi(\kappa),
\eeq where $\Phi(\kappa)$ may be considered at this point as a known
function of $\kappa$ ($\Phi(\kappa)$ is constructed from the initial
ansatz for the 4-point and 6-point functions). The equation above
can be easily solved, with the result \beq\label{solutionSchematic}
\Gamma(\kappa)=\left[ \Gamma(\kappa_0)+\int^\kappa_{\kappa_0}
\frac{d\kappa'}{\kappa'} \Phi(\kappa')
\left(\frac{\kappa_0}{\kappa'}\right)^\xi\right]\left(\frac{\kappa}{\kappa_0}\right)^\xi.\eeq
 Since, as we have just argued, $\Phi(\kappa)$ vanishes  as
$\kappa\to 0$,  the integral does not diverge faster than
$\kappa^{-\xi}$, and the small $\kappa$ behavior is governed by the
factor outside the brackets, that is by the solution of
eq.~(\ref{eqnapproxforgamma}), which guarantees that $\Gamma(0)=0$. 
This argument shows also that provided one solves consistently
eq.~(\ref{fulleqnforGamma}) small errors in the estimate of the
function $\Phi$ are damped by the factor $(\kappa/\kappa_0)^\xi$.  In
particular, the solution for the initial ansatz  is identical to
that written above, eq.~(\ref{solutionSchematic}), with $\Phi$
replaced by another function $\tilde\Phi$ not too different from $\Phi$ (we may ignore here the factor $F(\kappa)$ in eq.~(\ref{gamma40new}), which   complicates the
analysis, but in an inessential way. Accordingly we can also ignore the
contribution of the 6-point function in eq.~(\ref{4pointLOapp}). Then the only difference
between eqs.~(\ref{gamma40new}) and (\ref{4pointLOapp}) comes from
the approximation of the functions $J^{(3)}_3$ of eq.(\ref{4pointLOapp})
with the functions $\Theta$ in eq.~(\ref{gamma40new})). One then
expects  the two solutions corresponding to $\Phi$ and $\tilde\Phi$
to be close to each other. 

Let us however imagine that we apply the direct procedure to our
schematic equations, by caculating $\Gamma_{LO}$ from the analog of
eq.~(\ref{4pointLOapp}), namely \beq
\kappa\del_\kappa\Gamma_{LO}(\kappa)=\xi \tilde \Gamma(\kappa)+
\Phi(\kappa)=\kappa\del_\kappa \tilde
\Gamma(\kappa)+\Phi(\kappa)-\tilde\Phi(\kappa),  \eeq where we have
used the fact that $\kappa\del_\kappa
\tilde\Gamma(\kappa)=\xi\tilde\Gamma(\kappa)+\tilde\Phi(\kappa)$. We
would then obtain \beq \Gamma_{LO}(\kappa)-\tilde\Gamma(\kappa)
=\int_{\kappa_0}^\kappa\frac{d\kappa'}{\kappa'}\left(
\Phi(\kappa')-\tilde\Phi(\kappa')\right) \eeq where the integral is
convergent and yields a finite value for $\Gamma_{LO}(\kappa=0)$.
Thus, although the two solutions $\Gamma(\kappa)$ and
$\tilde\Gamma(\kappa)$ would differ little if $\Phi$ and
$\tilde\Phi$ differ little, the direct procedure which consists in
simply integrating  the r.h.s. of the flow equation fails to
reproduce the low $\kappa$ behavior.   This small $\kappa$
behavior can only be obtained if the feedback of the flow is
properly taken in the solution of the flow equation. But, as
revealed in the previous analysis, this needs to be done only in the
channel where the flow is not controlled by the external momenta,
i.e., the $s$-channel. We shall implement this improved strategy
for the LO in the next subsection.

\subsection{Improved approximation for $\Gamma^{(4)}_{12ll}(\kappa;p,-p,0,0)$ \label{gamma4LOi}}

As we have seen, the main problem with the direct procedure that we
have followed in the previous subsection comes from the $s$-channel
where the exceptional configuration of momenta does not contribute
to stabilize the flow when $\kappa\to 0$. In this subsection and the
following one, we shall present a more accurate treatment of this
particular situation,  making more precise the treatment presented
in the previous subsection.

To this aim, we consider first the case $q=0$ and replace
eq.~(\ref{4pointLOapp}) by: \beq \label{eqdif} \kappa\partial_\kappa
\Gamma^{(4)}_{12ll}(\kappa;p,-p,0,0)=
I_d^{(3)}(\kappa)\Gamma^{(4)}_{12ij}(\kappa;p,-p,0,0)\tilde\Gamma^{(4)}_{llij}(\kappa;0,0,0,0)
+ \Phi_{12}(\kappa;p), \eeq where we have isolated the contribution
of the $s$-channel, and grouped the other contributions into the
function $\Phi_{12}(\kappa;p)$:
\begin{eqnarray}
\label{defH}
\Phi_{12}(\kappa,p)&\equiv& J_d^{(3)}(\kappa;p)\tilde\Gamma^{(4)}_{1lij}(\kappa;p,0,0,-p)
\tilde\Gamma^{(4)}_{2lij}(\kappa;-p,0,0,p)
\nonumber \\
&+&J_d^{(3)}(\kappa;p)\tilde\Gamma^{(4)}_{1lij}(\kappa;p,0,0,-p)
\tilde\Gamma^{(4)}_{2lij}(\kappa;-p,0,0,p) \nonumber \\
&-&\frac{1}{2}I_d^{(2)}(\kappa)\tilde\Gamma^{(6)}_{12llmm}(\kappa;p,-p,0,0,0,0).
\end{eqnarray}
$\Phi_{12}(\kappa,p)$ is a known function which involves the initial
ansatz $\tilde\Gamma^{(4)}$ and $\tilde\Gamma^{(6)}$. In contrast,
$\Gamma^{(4)}_{12ll}(\kappa;p,-p,0,0)$ is taken to be the same
function in the r.h.s and the l.h.s. of eq.~(\ref{eqdif}).  That
is, eq.~(\ref{eqdif})  properly takes into account the feedback of
the flow in the $s$-channel on the solution of the flow equation. 
Note that for $p=0$ the solution of eq.~(\ref{eqdif}) (or
eq.~(\ref{4pointLOapp})) is the LPA solution. We can therefore
replace  in the r.h.s. of eq.~(\ref{eqdif})
$\tilde\Gamma^{(4)}_{ijkl}(\kappa;0,0,0,0)$ by the LPA value
\beq\label{GammaLPA} \tilde\Gamma^{(4)}_{ijkl}(\kappa;0,0,0,0)=
g(\kappa) (\delta_{ij} \delta_{kl} + \delta_{ik} \delta_{jl} +
\delta_{il} \delta_{jk}). \eeq Then eq.~(\ref{eqdif}) becomes \beq
\label{eqdif2} \kappa\partial_\kappa \Gamma^{(4)}(\kappa;p)= (N+2)
I_d^{(3)}(\kappa)g(\kappa)\Gamma^{(4)}(\kappa;p) + \Phi(\kappa;p),
\eeq where we used the definitions \beq \Gamma^{(4)}_{12ll}
(\kappa;p,-p,0,0) \equiv  \delta_{12} \Gamma^{(4)}(\kappa;p) ,
\hskip 1 cm  \Phi_{12}(\kappa;p)= \delta_{12} \Phi(\kappa;p), \eeq
and we omitted a  factor  $\delta_{12}$ in both sides of
eq.~(\ref{eqdif2}).

Eq.~(\ref{eqdif2}) is  a linear differential equation in $\kappa$,
with a non-homogenous term $\Phi$, depending on a parameter $p$. The
general solution of the homogenous equation  is: \beq \Gamma^{(4)}
(\kappa;p)= \Gamma^{(4)}(\kappa_0;p) \; e^{\int_{\kappa_0}^\kappa
\frac{d\kappa'}{\kappa'} \gamma(\kappa')}, \eeq where
$$
\gamma(\kappa) \equiv (N+2) I_d^{(3)}(\kappa)g(\kappa).
$$
Note that, as $\kappa\to 0$, $\gamma(\kappa) \sim \xi$ with $\xi>0$. The
value  of  $\kappa_0$ is to be chosen large enough for the LPA
solution to remain a good approximation at this value of $\kappa$
(for example, $\kappa_0 = 10 \; p$).

A particular solution of the total equation can be found in the
form: \beq \Gamma^{(4)} (\kappa;p)= \hat \Gamma^{(4)}(\kappa;p) \;
e^{\int_{\kappa_0}^\kappa \frac{d\kappa'}{\kappa'} \gamma(\kappa')}.
\eeq One gets \beq \hat \Gamma^{(4)} (\kappa;p) =
\int_{\kappa_0}^\kappa \frac{d\kappa''}{\kappa''} \Phi(\kappa'',p)
\; e^{-\int_{\kappa_0}^{\kappa''} \frac{d\kappa'}{\kappa'}
\gamma(\kappa')}+\Gamma^{(4)} (\kappa_0;p), \eeq so that the general
solution of equation (\ref{eqdif2}) can be written as
\beq\label{approx}
 \Gamma^{(4)} (\kappa;p) =\left[ \int_{\kappa_0}^\kappa \frac{d\kappa'}{\kappa'} \Phi(\kappa';p) \;
  e^{-\int_{\kappa_0}^{\kappa'} \frac{d\kappa''}{\kappa''}\gamma(\kappa'')} + \Gamma^{(4)}(\kappa_0;p)\right]
   \; e^{\int_{\kappa_0}^\kappa \frac{d\kappa'}{\kappa'} \gamma(\kappa')} .
\eeq This expression has the expected behavior: the last exponential
in the r.h.s. guarantees that $ \Gamma^{(4)} (\kappa,p) \to 0$ as a
power law  when $\kappa \to 0$.

The behavior of the solution is shown in fig.~\ref{gamma4_p=u} above
and compared  with the other expressions that we had for
$\Gamma^{(4)}$, for the particular values   $p=u$ and $p=u/10$.  One
can appreciate the effect of the improved procedure: the
corresponding curve follows the direct LO one down  to $\kappa \sim
p/10$, then it correctly goes to zero when $\kappa \to 0$  while the
direct LO one does not. This analysis confirms the importance of
keeping the feedback of the flow in the $s$-channel, that is, of
solving the flow equation in this channel, in order to get the
correct behavior at $\kappa \to 0$.

As suggested by fig.~\ref{gamma4_p=u}, the range of values of $\kappa$ where $\Gamma^{(4)}(\kappa;p,-p,0,0)$ obtained in the direct procedure exhibits its pathological behavior decreases as $p$ decreases. In fact, as $p\to 0$, $\Gamma^{(4)}(\kappa;p,-p,0,0)\to 0$ as $\kappa\to 0$ (since the 4-point function is then given by the LPA').
One can argue at this point that the unphysical behavior of
$\Gamma^{(4)}(\kappa)$ at small $\kappa$  has a moderate influence
on the calculation of the NLO result for the self-energy. Indeed, as
we shall see, it is mostly the region $\kappa \sim p$  that
contributes significantly to the flow of $\Sigma_{NLO}(p)$; in this
region, the estimates of the 4-point function obtained in the direct and improved LO
are almost identical (the difference with the initial
ansatz is also small). This is why we have used the direct procedure
in the first estimate of $\Sigma_{NLO}$ presented in
\cite{Blaizot:2004qa}. However, as we shall see in the next section,
the improved procedure  turns out to be much more accurate.
\subsection{Improved calculation of $\Gamma^{(4)}(\kappa;p,-p,q,-q)$}

The procedure described in the previous subsection only applies to
the 4-point function at $q=0$. It is only for this value of $q$ that
we can write eq.~(\ref{4pointLOapp}) as a closed equation: if $q\ne
0$, the 4-point function in the l.h.s. and those in the r.h.s. are
evaluated in different momentum configurations. However, we  note
that, in the calculation of $\Sigma_{NLO}(\kappa;p)$,  we need
$\Gamma^{(4)}(\kappa;p,-p,q,-q)$ only for $q<\kappa$, i.e., in a
range of values of $q$ that vanishes as $\kappa \to 0$.  In that
range,  we expect $\Gamma^{(4)}(\kappa;p,-p,q,-q) $ to differ very
little from  $\Gamma^{(4)}(\kappa;p,-p,0,0)$. Proceeding as in the
previous subsection, we single out the $s$-channel and rewrite
eq.~(\ref{4pointLOapp}) as \beq\label{equadifq}
\kappa\partial_\kappa \Gamma^{(4)}_{12ll}(\kappa;p,-p,q,-q)=
I_d^{(3)}(\kappa)\Gamma^{(4)}_{12ij}(\kappa;p,-p,0,0)\Gamma^{(4)}_{llij}(\kappa;q,-q,0,0)
+ \Phi_{12}(\kappa,p,q) \eeq where, now
\begin{eqnarray}
\Phi_{12}(\kappa,p,q)&=&J_d^{(3)}(\kappa;p+q)\tilde\Gamma^{(4)}_{1lij}(\kappa;p,q,0,-p-q)
\tilde\Gamma^{(4)}_{2lij}(\kappa;-p,-q,0,p+q)
\nonumber \\
\hspace{.5cm}&+&J_d^{(3)}(\kappa;p-q)\tilde\Gamma^{(4)}_{1lij}(\kappa;p,-q,0,-p+q)
\tilde\Gamma^{(4)}_{2lij}(\kappa;-p,q,0,p-q) \nonumber \\
\hspace{.5cm}&-&\frac{1}{2}I_d^{(2)}(\kappa)\tilde\Gamma^{(6)}_{12llmm}(\kappa;p,-p,q,-q,0,0)
\\ \nonumber & \equiv& \delta_{12} \Phi(\kappa;p,q).
\end{eqnarray}
In eq.~(\ref{equadifq}), we use the fact that $q<\kappa$ to replace
$\Gamma^{(4)}_{llij}(\kappa;q,-q,0,0)$ by the initial ansatz (we
have seen earlier that this is a good approximation): \beq
\Gamma^{(4)}_{llij}(\kappa;q,-q,0,0) \rightarrow (N+2) \delta_{ij} g(\kappa).
\eeq As a result, all the $q$-dependence is now in the function
$\Phi(\kappa;p,q)$.  For the other factor in the r.h.s. of
eq.~(\ref{equadifq}), $\Gamma^{(4)}_{llij}(\kappa;p,-p,0,0)$, we use
the improved solution obtained in the previous subsection, which we
shall denote as $\Gamma^{(4)}_{12ll}(\kappa;p,-p,0,0) =
\delta_{12}\bar \Gamma^4(\kappa,p)$. The $4$-point function
$\Gamma^{(4)}_{12ll}(\kappa;p,-p,q,-q) $ can then be obtained by
simply integrating the r.h.s. of eq.~(\ref{equadifq}) between
$\Lambda$ and $\kappa$ and adding the initial value (see eq. (12)).
One then gets
\begin{eqnarray}\label{appq}
\Gamma^{(4)}_{12ll}(\kappa;p,-p,q,-q) &= &\delta_{12}(N+2)
\frac{u}{3}  \\ \nonumber &+&\delta_{12} \int_\Lambda^\kappa
\frac{d\kappa'}{\kappa'} \left[ \gamma(\kappa') \bar
\Gamma^4(\kappa';p) + \Phi(\kappa';p,q) \right],
\end{eqnarray}
where
\begin{eqnarray}
\bar \Gamma^4(\kappa;p) = \left\{
\begin{array}{lc}
(N+2) g(\kappa)&\,\,\,\, \mathrm{if} \kappa> \kappa_0 \\
\int_{\kappa_0}^{\kappa} \frac{d\kappa'}{\kappa'}
\Phi(\kappa';p) \; e^{\int_{\kappa'}^{\kappa} \frac{d\kappa''}{\kappa''} \gamma(\kappa'')}
+ (N+2) g(\kappa_0) \; e^{\int_{\kappa_0}^{\kappa} d\kappa' \gamma(\kappa')}
&\,\,\,\,\mathrm{if} \kappa \leq \kappa_0 \, .\\
\end{array}
\right.
\nonumber\\
\end{eqnarray}

\subsection{The dependence on $\alpha$}
\label{gamma4_alpha}

In paper I we discussed the dependence of $\Sigma_{LO}$ on the
parameter $\alpha$ that we introduced in one of the approximations
(approximation ${\cal A}_2$) used to construct the initial ansatz
for the 4-point function (see eq.~(\ref{gamma40new})). Recall that
this approximation consists in replacing  propagators such as  $
G(p+q)$ in the r.h.s. of the flow equations by
$G_{LPA'}(q)\Theta(\kappa^2-\alpha^2 p^2)$. We found in paper I that
$\Sigma_{LO}(p;\alpha)$ obeys an approximate scaling law,
$\Sigma_{LO}(p;\alpha)\simeq \hat \Sigma(\alpha p)$. Here, we
discuss the $\alpha$-dependence of the LO expression of the 4-point
function. For simplicity, we shall discuss the dependence of the
expression obtained using the direct procedure
(sect.~\ref{gamma4LOd}). The $\alpha$-dependence of that obtained with the
improved procedure is essentially identical.

\begin{figure}[t!]
\begin{center}
\includegraphics*[width=10 cm,angle=-90]{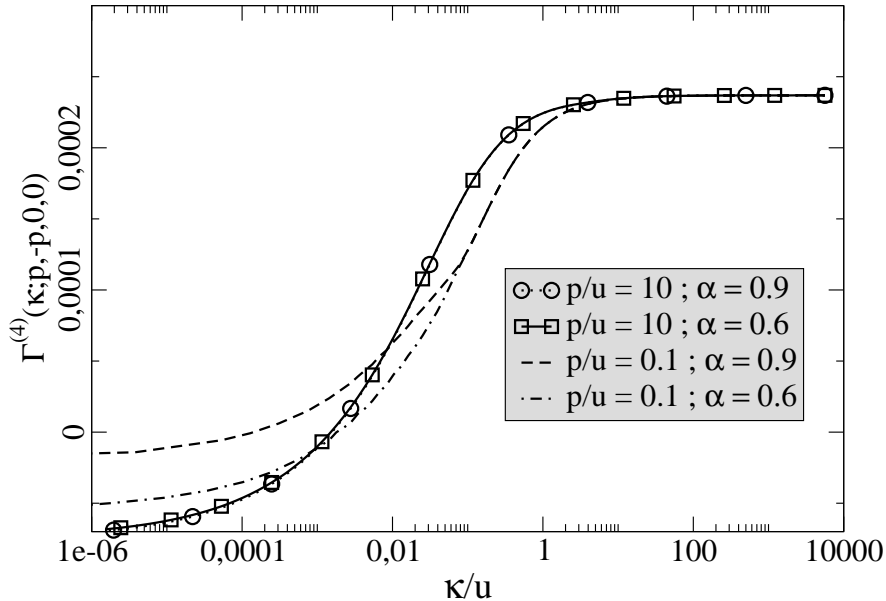}
\end{center}
\caption{\label{gamma4_alpha_k} The function $\Gamma^{(4)}
(\kappa,p,-p,0,0)$ (in units of $\Lambda$) for the values $p=10 u$
and $p=u/10$ as a function of $\kappa/u$, for  $\alpha =0.6$ and
$\alpha=0.9$. The $\alpha$-dependence is largest when $p\simeq
\kappa_c\simeq u/10$. When $p=10 u\gg \kappa_c$, $\Gamma^{(4)}$ is
independent of $\alpha$.  }
\end{figure}

 To study the $\alpha$-dependence of the 4-point function, it is convenient to separate
$\Gamma^{(4)}_{12ll}(\kappa;p,-p,q,-q)$ into three contributions:
that of  the $s$-channel (see fig.~\ref{ti}), denoted by
$\Gamma^{(4)[s]}$; the sum of the contributions of the $t$ and
$u$-channels (see fig.~\ref{sunset}), denoted by
$\Gamma^{(4)[t+u]}$; finally the contribution of the 6-point
function (see fig.~\ref{octupus}),  denoted as $\Gamma^{(4)[6]} $. That is, $\Gamma^{(4)[i]}$
denotes the contribution to the 4-point function obtained when only
the channel $i$ is included in the calculation of $\Gamma^{(4)}$
according to eq.~(\ref{4pointLOapp}).
There is an important difference between $\Gamma^{(4)[s]}$ and
$\Gamma^{(4)[6]} $ on one side, and $\Gamma^{(4)[t+u]}$ on the other
side:  while $p$ flows through  the loop in $\Gamma^{(4)[t+u]}$ (see
fig.~\ref{sunset}),  in the other two cases it does not, so that the
$p$-dependence of $\Gamma^{(4)[s]}$ and $\Gamma^{(4)[6]} $ comes
entirely from the vertices. The latter are  those of the initial
ansatz $\tilde\Gamma^{(4)}$ and $\tilde\Gamma^{(6)}$, which depend
on $p$ and $\alpha$   approximately  only through the product $p
\alpha$ (see paper I, sect. III and app.~\ref{sec:gamma6}). On the
other hand, the $p$-dependence of $\Gamma^{(4)[t+u]}$ is mainly due
to the explicit $p$-dependence of the loop in fig. \ref{sunset},
which is given by the ($\alpha$-independent) function
$J_d^{(3)}(\kappa;p)$: Since $J_d^{(3)}(\kappa;p)$ is only
important when $\kappa \simge p$ (see fig.~9  in paper I), the
contribution of the $t+u$ channels is non zero only in a region
where the vertices in fig.~\ref{sunset} are essentially the ($p$ and
$\alpha$ independent) LPA' ones (see e.g. fig.~\ref{gamma4_p=u}).
Then, one expects $\Gamma^{(4)[t+u]}_{12ll}(\kappa;p,-p,q,-q)$ to be
almost independent of $\alpha$.

Fig.~\ref{gamma4_alpha_k} shows that the total 4-point function
$\Gamma^{(4)} (\kappa,p,-p,0,0)$ is in fact almost independent of
$\alpha$ when $p=10 u$, which indicates that  the contributions of
the $t$ and $u$ channels  dominate for this value of $p$.  The
same holds for all values of $p$ much larger of much smaller than
$u$. Only in the intermediate momentum region $p \sim \kappa_c \sim
u/10$, is the variation with $\alpha$ important, which reflects the
fact that in this intermediate range of momenta, the contributions
of $\Gamma^{(4)[s]}$ and $\Gamma^{(4)[6]}$  are of the same
order of magnitude as that of $\Gamma^{(4)[t+u]}$,  as we shall
verify later (see e.g. fig.~7 below, and the related discussion
concerning the $\alpha$-dependence of $\Sigma_{NLO}(p)$).

\section{The self-energy and $\Delta\langle \varphi^2\rangle$ at NLO}

\label{sec-selfNLO}

We have now all the ingredients to calculate the self-energy at
next-to-leading order. Recall that the physical self-energy at
criticality  is given by (see eq.~(I.108)) \beq \label{selfLO1b}
\delta_{12}\Sigma(p)
=\frac{1}{2}\int_0^{\Lambda}d\kappa'\int\frac{d^dq}{(2\pi)^d}
G^2(\kappa';q)\partial_{\kappa'}R_{\kappa'}(q)
\left(\Gamma^{(4)}_{12ll}(\kappa';p,-p,q,-q)-\Gamma^{(4)}_{12ll}(\kappa';0,0,q,-q)
\right). \nonumber \\
\eeq
In order to get $\Sigma_{NLO}$, one needs to insert in the r.h.s. of eq.~(\ref{selfLO1b}) the
 LO expression for the 4-point function $\Gamma^{(4)}_{12ll}(\kappa;p,-p,q,-q)$ which
has been calculated in the previous  section, and  the LO propagator
given by: \beq
G^{-1}(\kappa;q)=q^2+\Sigma_{LO}(\kappa;q)+R_\kappa(q), \eeq with
$\Sigma_{LO}(\kappa;q)$  the LO expression of the self-energy, given
by eq.~(I.111).

\subsection{Self-energy $\Sigma_{NLO}$}

\begin{figure}[t!]
\begin{center}
\includegraphics*[width=12 cm]{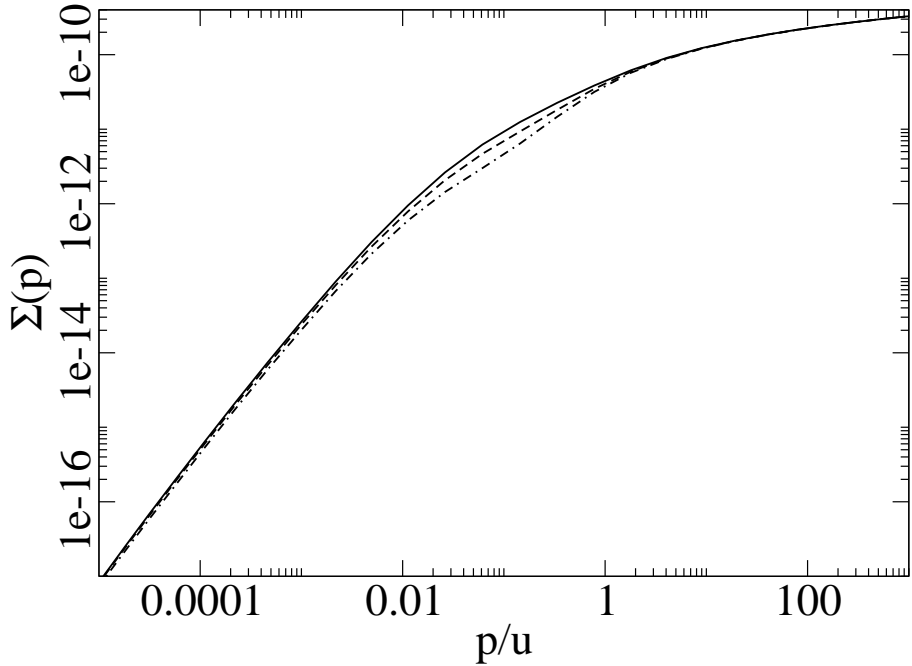}
\end{center}
\caption{\label{selfNLO_alpha} Self-energy at NLO (in units of $\Lambda^2$) as a function of
$p/u$
 for various values of $\alpha$: $\alpha=0.6$ (full line),
$\alpha=0.75$ (dashed line) and $\alpha=0.9$ (dot-dashed line). }
\end{figure}

In this subsection, we present numerical results obtained for
$d=3$ and $N=2$ in order to illustrate the main features of the
self-energy. We shall use here the LO estimate of the 4-point
function derived in the direct procedure. In the next
subsection, we shall discuss results obtained with the improved
procedure; we shall also   present  results for $N\ne 2$.

Fig.~\ref{selfNLO_alpha} displays the self-energy at NLO for various
values of $\alpha$. It is to be compared with fig.~12 in paper I,
for the LO behavior of $\Sigma(p)$. Note, however, that in paper I
$\Sigma_{LO}$ is plotted as a function of $(\alpha p/u)$ in order to
exhibit its approximate scaling property
$\Sigma_{LO}(\alpha;p)\approx \hat\Sigma(\alpha p)$. Here
$\Sigma_{NLO}$ is plotted as a function of $(p/u)$; as can be seen
on fig.~\ref{selfNLO_alpha}, $\Sigma_{NLO}$ depends much less on
$\alpha$ than $\Sigma_{LO}$. In fact,  both the IR and the UV
regimes are nearly independent of $\alpha$. It is only in the
intermediate momentum range ($p\sim \kappa_c$)  that  $\Sigma_{NLO}$ exhibits some
dependence on $\alpha$.

At high momentum, one expects $\Sigma(p)$ to be given by
perturbation theory, that is, one expects $\Sigma(p)\sim \ln(p/u)$.
Recall that in paper I, we found that $\Sigma_{LO}(p)$ indeed  behaves in
this way, but the coefficient in front of the logarithm
differed from that of perturbation theory by 7\%.   As we have discussed in paper I,  perturbation theory is recovered exactly at high momenta when one performs iterations. In particular, the 2-loop result is exactly reproduced at NLO. We have verified
 that the coefficient in front of the logarithm in $\Sigma_{NLO}(p)$ at large $p$, is
correctly obtained, to within the numerical accuracy with which it
can be determined (about 0.5\%).

In the IR regime, the power law behavior already reproduced  in  LO, $p^2+\Sigma_{LO}(p)\sim p^{2-\eta}$,
 is essentially not modified:  however the numerical calculation is more involved in NLO, leading to a loss of accuracy that  prevents us to determine
the value of $\eta$ with any useful precision.

 As was the case in LO, the $\alpha$-dependence of
$\Sigma_{NLO}$ is intimately connected with its momentum dependence.
Following the analysis that we did in sub-section \ref{gamma4_alpha}
to understand the variation of $\Gamma^{(4)}$ with $\alpha$, we
split $\Sigma_{NLO}$ into three separate contributions: we define
$\Sigma^{[s]}$, $\Sigma^{[t+u]}$ and $\Sigma^{[6]}$ as the
contributions obtained, respectively,  when only $\Gamma^{(4)[s]}$,
$\Gamma^{(4)[t+u]}$ and $\Gamma^{(4)[6]}$ are included  in the
r.h.s. of eq.~(\ref{selfLO1b}).  The properties of the LO 4-point
function discussed in the previous section imply that
$\Sigma^{[s]}(\alpha;p)\approx\bar \Sigma^{[s]}(\alpha p)$,
$\Sigma^{[6]}(\alpha;p)\approx\bar \Sigma^{[6]}(\alpha p)$ (i.e.,
the same dependence on $\alpha$ as $\Sigma_{LO}$ (see paper I,
sect.~IV A)), while $\Sigma^{[t+u]}$ is expected to be roughly
independent of $\alpha$. These properties are well verified in our
numerical calculations. 

\begin{figure}[t!]
\begin{center}
\includegraphics*[width=12 cm]{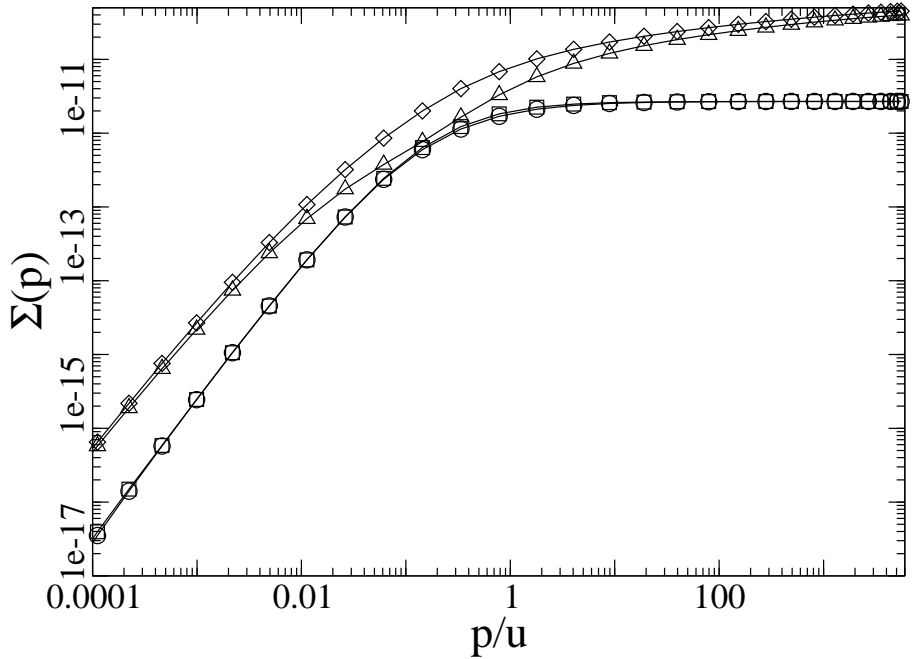}
\end{center}
\caption{\label{self-contr} Self-energy $\Sigma_{NLO}(p)$ (in units
of $\Lambda^2$) as a function of $p/u$   (triangle) at $\alpha=0.83$
and its three partial contributions $\Sigma^{[t+u]}$ (diamond),
$-\Sigma^{[s]}$ (circle) and $-\Sigma^{[6]}$ (square).}
\end{figure}

The three contributions of the self-energy  are shown in
fig.~\ref{self-contr} together with their sum $\Sigma_{NLO}$. While $\Sigma^{[t+u]}$ is
positive, the two other contributions  are negative.  The latter
property can be understood as follows: In calculating, say,
$\Sigma^{[s]}$, one puts in eq.~(\ref{selfLO1b}) only
$\Gamma^{(4)[s]}$ which, in turn, is calculated with only the first
term of eq.~(\ref{4pointLOapp}). In the latter, the flow is
evaluated with the initial ansatz $\tilde\Gamma^{(4)}$ which, as can
be seen in sect. III C of paper I, verifies, in all regions of
momenta, $\tilde\Gamma^{(4)}(\kappa;p,-p,q,-q)
> \tilde\Gamma^{(4)}(\kappa;0,0,q,-q)>0$. Since  the integration over $\kappa'$ that
is needed to obtain $\Gamma^{(4)[s]}(\kappa)$ from
eq.~(\ref{4pointLOapp}) runs from $\Lambda$ to $\kappa$, we have
$\Gamma^{(4)[s]}(\kappa;p,-p,q,-q)<
\Gamma^{(4)[s]}(\kappa;0,0,q,-q)$ so that the integrand in
eq.~(\ref{selfLO1b}) is negative, yielding eventually
$\Sigma^{[s]}<0$. A similar analysis can be done for $\Sigma^{[6]}$.
 Fig.~\ref{self-contr} shows that, as was the case for the 4-point
function discussed in the previous section,  the $t$ and $u$
channels dominate, except in the intermediate momentum region
($p\sim \kappa_c$) where the contributions of the three channels are
of the same order of magnitude.  This, together with the $\alpha$-dependence
of $\Sigma^{[s]}$, $\Sigma^{[6]}$ and $\Sigma^{[t+u]}$ recalled in the previous paragraph, explains the behavior seen in fig.~\ref{selfNLO_alpha}. 
Fig.~\ref{self-contr} also reveals an interesting feature of the
present approximation, for which we have no simple explanation: to
within numerical accuracy, $\Sigma^{[s]}$ and $\Sigma^{[6]}$ appear
indistinguishable. This property remains true for other values of
$N$.

\subsection{Calculation of $\Delta\langle \varphi^2\rangle$}

We turn now to the calculation of the changes of the fluctuations of
the field caused by the interactions: \beq\label{integralc}
\frac{\Delta\langle \phi_i^2\rangle}{N}= \int\frac{\d^3
p}{(2\pi)^3}\,\left( \frac{1}{p^2+\Sigma(p)}-\frac{1}{p^2}\right)=
\frac{1}{2\pi^2}\int\frac{dp}{p}\left[
\frac{p^3}{p^2+\Sigma(p)}-p\right]. \eeq As recalled in the
introduction, and more thoroughly in paper I, this quantity is very
sensitive to the intermediate momentum region and constitutes a
stringent test of the calculation. In the following, we shall refer
to the quantity in square brakets in eq.~(\ref{integralc}),
multiplied by $1/(2\pi^2)$, as to the integrand. Note that in the
range of momenta where this integrand  is significant (see e.g. fig.~\ref{cont-tad-sun-oct} below), $\Sigma(p)
\ll p^2$, so that the integrand can be well approximated by
$-\Sigma(p)/(2\pi^2\, p)$.

The results for $\Delta\langle \phi_i^2\rangle$ will be discussed in
terms of the parameter \beq c\,\equiv -\frac{256
\pi^3}{\left(\zeta(3/2)\right)^{4/3}} \,\frac{\Delta\langle\varphi_i^2\rangle}{Nu}.\eeq The shift, caused by weak interactions, in the
temperature of the Bose-Einstein transition of a dilute gas is directly proportional to $c$ \cite{club,bigbec}.

\begin{figure}[t!]
\begin{center}
\includegraphics*[width=10 cm]{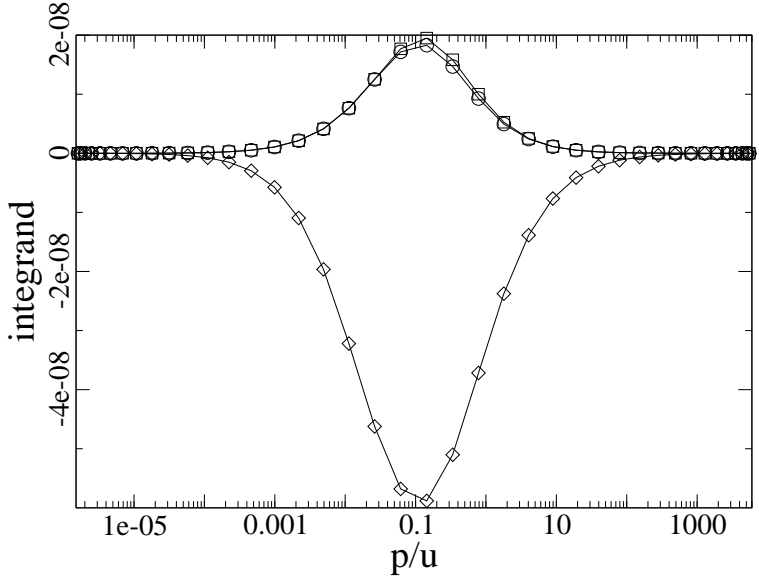}
\end{center}
\caption{ \label{cont-tad-sun-oct} The three curves represent the
integrand of eq.~(\ref{integralc}) (in units of $\Lambda$) calculated with only
$\Sigma^{[t+u]}$ (diamonds), $\Sigma^{[s]}$ (circle) and
$\Sigma^{[6]}$ (squares) contributions to the self-energy at NLO,
respectively, as a function of $p/u$ (the points shown are those
needed in the numerical calculation of the integral in
eq.~(\ref{integralc})). The plots correspond to $\alpha=0.83$ and $N=2$.}
\end{figure}

As we have seen in fig.~\ref{selfNLO_alpha}, $\Sigma_{NLO}(p)$ is
independent of the parameter $\alpha$ both at high and low momenta.
However, in the crossover region ($p\sim \kappa_c$)  which determines the value of $c$, $\Sigma_{NLO}(p)$
still depends on the value of $\alpha$.  It follows
that the value of the coefficient $c$ calculated with $\Sigma_{NLO}(p)$
still depends on $\alpha$. To understand better the $\alpha$-dependence of the NLO predictions,  one can write
the coefficient $c$  as the sum of three expressions, each of them
containing only one of the three parts of the self-energy. The three
contributions to the integrand yielding $c$ are displayed together
in  fig.~\ref{cont-tad-sun-oct}.
Because of the approximate scaling   discussed  above, both
$\Sigma^{[s]}$ and $\Sigma^{[6]}$ contribute to $c$ with a term
proportional to $\alpha$, while  the contribution of
$\Sigma^{[t+u]}$  is essentially independent of $\alpha$. This afine
behavior of the NLO result for $c$ is indeed observed in
fig.~\ref{c-LO-NLO} below. The negative slope is due to the negative sign
of $\Sigma^{(4)[s]}$ and $\Sigma^{(4)[6]}$.

The $\alpha$-dependence remains a source of uncertainty in the calculation of $c$. As can be seen on fig.~\ref{c-LO-NLO}, when we move from the LO calculation to the  direct NLO, to the  improved NLO, the dependence on $\alpha$ decreases, and so does the corresponding uncertainty in the calculated value of $c$. We  regard the variation in the value of $c$ when $\alpha$ runs form $.6$ to $.9$ as a large  overestimate of the uncertainty related to the choice of $\alpha$.  In fact we can eliminate much of this uncertainty by  following a procedure suggested by the results plotted in fig.~\ref{c-LO-NLO}: since the curves representing $c$ as a function of $\alpha$ have opposite slopes at LO and NLO, one can invoke a principle of fastest apparent
convergence to choose as best estimate that given by the value of $\alpha$ for which the two curves cross: At this point indeed, the NLO correction vanishes. One thus obtains
the value $c=1.44$ (the crossing point being at $\alpha= 0.83)$ when we use the direct LO expression of the 4-point function, and $c=1.30$ with the improved LO (corresponding to $\alpha=0.77$). The improved NLO calculation is thus in remarkable agreement with the lattice data:  $1.32\pm 0.02$ \cite{latt2} and $1.29\pm 0.05 $ \cite{latt1}. 

\begin{figure}[t!]
\begin{center}
\includegraphics*[width=10 cm,angle=-90]{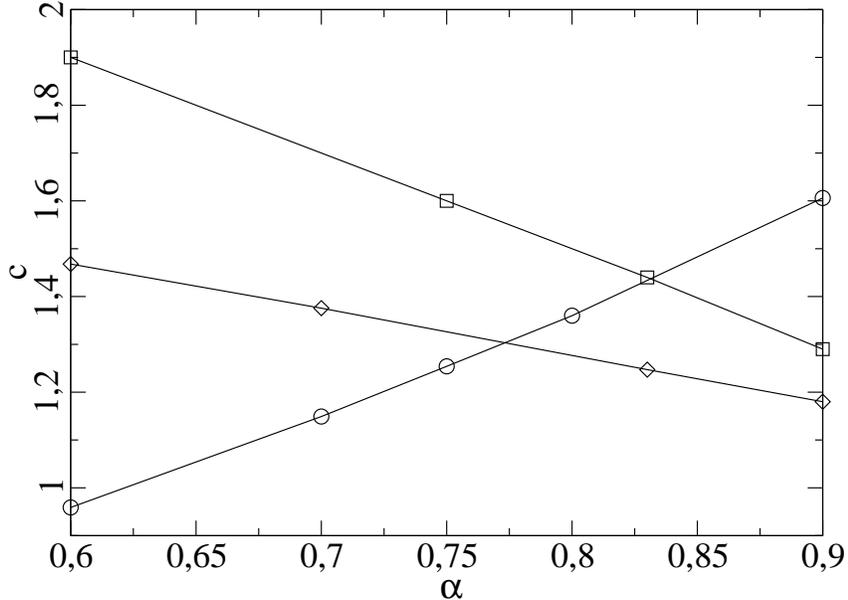}
\end{center}
\caption{\label{c-LO-NLO} The coefficient $c$
 as a function of $\alpha$ for $N=2$: LO (circles) from paper I, NLO direct (squares), NLO improved (diamonds).}
\end{figure}

\begin{figure}[t!]
\begin{center}
\includegraphics*[width=12 cm,angle=-90]{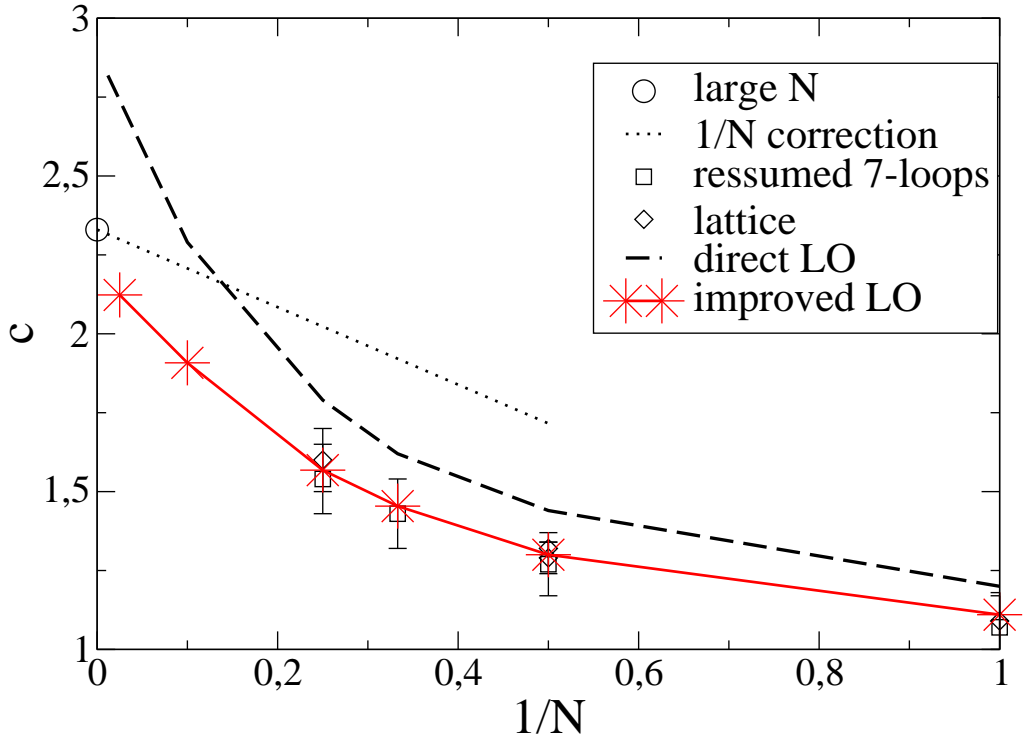}
\end{center}
\caption{\label{coefc_N} The coefficient $c$ (obtained with the
fastest apparent convergence procedure) as a
function of $1/N$. Our NLO results, are compared with
results obtained with other methods: lattice \cite{latt2,latt1,latt3}
(diamonds) and 7-loops perturbation theory \cite{Kastening:2003iu}
(squares), all of them with their corresponding error bars, together
with the $N\to\infty$ result \cite{BigN} (circle), and the extrapolation following the $1/N$ correction calculated in ref.~\cite{Arnold:2000ef}.}
\end{figure}

 We have also repeated our calculation for other values of $N$
for which results have been obtained with other techniques, either
the lattice technique \cite{latt2,latt1,latt3}, or variationally
improved 7-loops perturtbative calculations \cite{Kastening:2003iu}.
These results are summarized in table I and fig.~\ref{coefc_N}
(other optimized perturbative calculations have also been recently
performed, and are in agreement with those quoted here; see
\cite{souza,Kneur04}).
 For small  values ($N
\simle 10$), our results fulfill all the numerical tests that we
have described in this paper. For $N\le 4$, where we can compare
with other results, the values of $c$ obtained with the present
improved NLO calculation are in excellent agreement with those
obtained from lattice and 7-loops calculations\footnote{As mentioned
in \cite{Blaizot:2004qa} we regard the good agreement with lattice
results obtained in  the application of the NPRG to the calculation
of $\Delta T_c$ presented in  ref. ~ \cite{Ledowski03} as largely
accidental.  Indeed several approximations are done in ref. ~
\cite{Ledowski03} that constitute large sources of uncertainties:
only the $t$ and $u$ channel contributions are kept in the flow
equation  for the 4-point function, and the role of the exceptional
momenta is not recognized; vertices are approximated by essentially
the LPA' vertices, and the solution of the LPA' is also very
approximate. Finally the use of a sharp-cut off is known to be
problematic in conjunction with the derivative expansion. }.

What happens at large values of $N$ deserves a special discussion.
As seen in fig.~\ref{coefc_N} the curve showing the improved leading
order results extrapolates  when $N\to \infty$ to a value that is
about $4\%$ below the known exact result \cite{BigN}. A direct
calculation at very large values of $N$ is difficult in the present
approach for numerical reasons: since the coefficient $c$ represents
in effect an order $1/N$ correction (see \cite{BigN}), it is
necessary to insure the cancellation of  the large, order $N$,
contributions to the self-energy, in order to extract the value of
$c$. This is numerically demanding when $N\simge 100$.
Fig.~\ref{coefc_N} also reveals an intriguing feature: there seems
to be no natural way to reconcile the present results, and for this
matter the results from lattice calculations or 7-loop calculations,
with the calculation of the $1/N$ correction presented in
ref.~\cite{Arnold:2000ef}: the dependence in $1/N$ of our results,
be they obtained from the direct LO or the improved LO, appear to be
incompatible with the slope predicted by the $1/N$ expansion.

\begin{table}
\begin{tabular}{|c|c|c|c|c|c|c|c|}
\hline $c$ & $N=1$ & $N=2$ & $N=3$ & $N=4$ & $N=10$ & $N=40$ &
$N=\infty$
\\ \hline
lattice \cite{latt2} & & $1.32 \pm 0.02$ & & & & & \\
lattice \cite{latt1} & & $1.29 \pm 0.05$& & & & & \\
lattice \cite{latt3} & $1.09\pm 0.09$ & & & $1.60\pm 0.10$ & & & \\
7-loops \cite{Kastening:2003iu} & $1.07\pm 0.10$ & $ 1.27 \pm 0.10$ & $1.43\pm 0.11$ & $1.54\pm 0.11$ & & & \\
large $N$ \cite{BigN} & & & & & & &$c=2.33$ \\
this work  & 1.11 & 1.30  & 1.45 & 1.57 & 1.91 & 2.12 & \\
\hline
\end{tabular}
\caption{Summary of available results for the coefficient $c$. The
last line contains the results obtained in this work by using the
improved LO approximation for the 4-point function. }
\end{table}

\section{Summary and outlook}
\label{conclusions}

The quality of the results that we have obtained for the parameter $c$ that characterizes the shift in the transition temperature of the weakly repulsive Bose gas is encouraging. It demonstrates that the method that we have developed in order to solve the Non Perturbative Renormalisation Group is capable indeed to yield the full momentum dependence of the $n$-point functions, in physical regimes where other approximation schemes are limited.

It is of course difficult to quantify the size of the theoretical
uncertainties in this approach. We have commented in the previous section about the source of uncertainty related to the choice of the parameter $\alpha$. A  better  estimate of the  accuracy of the whole approximation scheme 
would be to perform one more iteration.  However the numerical effort
involved in the calculation of the relevant multidimensional
integrals is non negligible, and furthermore the approximate
equations that need to be solved in order to get the initial ansatz
for the $n$-point functions become increasingly complicated (see for
instance app.~\ref{sec:gamma6}). Thus it seems unrealistic to
imagine doing a further iteration, going say to the
next-to-next-to-leading order. What remains then, as a measure of
the quality of the approximation scheme,  is the direct comparison
with results obtained by other, reliable, non perturbative methods,
such as lattice calculations.

We have however started exploring an alternative approach which may have, among other features, the capability of yielding estimates for theoretical uncertainties. This  approach builds  on the  present approximation scheme, but brings to it conceptual simplification. As we have seen when discussing results, both in paper I and in the present paper, the approximation ${\cal A}_1$ introduced in paper I is accurate in most situations encountered when solving the flow equations. This approximation assumes that the vertices in the r.h.s. of the flow equation are smooth functions of the momenta and exploits the fact that the loop momentum is bounded to neglect some of the momentum dependence. Once this is done, as shown in \cite{PLB}, one can relate directly the higher $n$-point functions that arise in the r.h.s. of the flow equations to derivatives of the $n$-point function whose flow is being studied  with respect to a constant background field. One then obtain closed equations. This method allows us to bypass both the approximation ${\cal A}_3$ needed for $n$-point functions of high order, and also the approximation  ${\cal A}_2$ used to implement the decoupling of high momenta in the propagators. The price to pay is that the new equations are not only differential equations in the variable $\kappa$, but also partial differential equations with respect to the background field. However, they can be solved numerically, as will be demonstrated in a forthcoming publication \cite{BMWn}.

\appendix

\section{The initial ansatz for $\Gamma_{12llmm}^{(6)}(\kappa;p,-p,0,0,0,0)$ \label{sec:gamma6}
}

In order to obtain the explicit expression of
the initial ansatz for $\Gamma^{(6)}$  we
follow the same steps as in paper I, sect. III,  when constructing the initial
ansatz for $\Gamma^{(4)}$: We  use the three approximations
${\cal A}_1$, ${\cal A}_2$ and ${\cal A}_3$ to get an approximate equation for the flow of
$\Gamma^{(6)}$, which we then solve semi-analytically.

The exact flow equation for $\Gamma^{(6)}$ can be written in the following form, exhibiting  three  kinds of contributions:
\begin{eqnarray}
\label{eqcompletegamma62} \kappa\partial_\kappa
\Gamma_{123456}^{(6)}&=&\Tr G^2 \kappa\partial_\kappa R_\kappa
\left\{ - \Gamma_{i12j}^{(4)}G
\Gamma_{j34k}^{(4)}G\Gamma_{k56i}^{(4)} + \mathrm { 44\,
permutations}  \right.
\nonumber \\
&+& \left. \Gamma_{i1234j}^{(6)}G \Gamma_{j56i}^{(4)}+
\mathrm { 14\, permutations}-\frac{1}{2}\Gamma_{123456ii}^{(8)} \right\} .
\end{eqnarray}

We start by implementing approximations ${\cal A}_1$ and ${\cal A}_2$. To keep
the discussion simple, we  do so explicitly only for some typical
terms of eq.~(\ref{eqcompletegamma62}). Take for example the following
contribution involving the product of three $\Gamma^{(4)}$:
\begin{eqnarray}
&& - \Tr G^2 \kappa\partial_\kappa R_\kappa
 \Gamma_{i12j}^{(4)}G \Gamma_{j34k}^{(4)}G\Gamma_{k56i}^{(4)} =
- \int \frac{d^dq}{(2\pi)^d}  \kappa\partial_\kappa R_\kappa (q^2)
G(q^2)
\Gamma_{i12j}^{(4)}(q,p_1,p_2,-q-p_1-p_2)  \nonumber \\
&& \times\;\;  G((q+p_1+p_2)^2)
\Gamma_{j34k}^{(4)}(q+p_1+p_2,p_3,p_4,-q+p_5+p_6)\nonumber \\
&& \times\;\; G((q-p_5-p_6)^2)
\Gamma_{k56i}^{(4)}(q-p_5-p_6,p_5,p_6,-q) .
\end{eqnarray}
After setting the external momenta to their values
($p_3=p_4=p_5=p_6=0, p_2=-p_1$), imposing $q=0$ in the vertices (approximation
${\cal A}_1$) and replacing $G(q+p)$ by $G_{LPA'},
\Theta(1-\alpha^2p^2/q^2)$ (approximation ${\cal A}_2$), one gets:
\begin{eqnarray}
\label{gamma6t1} && - \Tr G^2  \kappa\partial_\kappa R_\kappa
 \Gamma_{i12j}^{(4)}G \Gamma_{j34k}^{(4)}G\Gamma_{k56i}^{(4)} = \nonumber \\
&& - I_d^{(4)}(\kappa)
\Gamma_{12ij}^{(4)}(p,-p,0,0)
\Gamma_{34jk}^{(4)}(0,0,0,0)
\Gamma_{56ki}^{(4)}(0,0,0,0) ,
\end{eqnarray}
where we have also made use of the symmetry of the bosonic
$n$-point functions. A similar contribution corresponding to a
different permutation reads:
\begin{eqnarray}
\label{gamma6t1b} && - \Tr G^2  \kappa\partial_\kappa R_\kappa
 \Gamma_{i13j}^{(4)}G \Gamma_{j24k}^{(4)}G\Gamma_{k56i}^{(4)} = \nonumber \\
&& - I_d^{(4)}(\kappa) \Theta(1-\alpha^2p^2/\kappa^2)
\Gamma_{1ji3}^{(4)}(p,-p,0,0) \Gamma_{j24k}^{(4)}(p,-p,0,0)
\Gamma_{56ki}^{(4)}(0,0,0,0) .
\end{eqnarray}
All 45 permutations containing three $\Gamma^{(4)}$ reduce either
to the forms (\ref{gamma6t1}) or   (\ref{gamma6t1b}): those where
both the external legs carrying the non-zero momenta ($p$ and
$-p$) belong to the same $\Gamma^{(4)}$ are of the type
(\ref{gamma6t1}), whereas those where the two legs with  non-zero
 momenta belong to two different $\Gamma^{(4)}$ are of the
 type
(\ref{gamma6t1b}).

The second kind of contributions is that which
involve  one $\Gamma^{(4)}$ and one $\Gamma^{(6)}$. Among the 15
contributions of this kind, we have three different cases,
depending on how the two external legs carrying non  zero momenta
are attached: both on $\Gamma^{(4)}$, both on $\Gamma^{(6)}$, one
on $\Gamma^{(4)}$ and the other on $\Gamma^{(6)}$. We explicitly
write here one contribution of the latter type: \beq
\label{gamma6t2}
 \Tr G^2  \kappa\partial_\kappa R_\kappa
 \Gamma_{1345ij}^{(6)}G \Gamma_{26ji}^{(4)} =
I_d^{(3)}(\kappa) \Theta (1-\alpha^2 p^2/\kappa^2)
\Gamma_{1j4513}^{(6)}(p,-p,0,0,0,0) \Gamma_{j26i}^{(4)}(p,-p,0,0)
. \nonumber\\\eeq

Finally, the third  type of contributions is that which involves
$\Gamma^{(8)}$: \beq \label{gamma6t3}
 -\frac{1}{2} \Tr G^2  \kappa\partial_\kappa R_\kappa
 \Gamma_{123456ij}^{(8)} =
-\frac{1}{2} I_d^{(2)}(\kappa)
\Gamma_{123456ii}^{(8)}(p,-p,0,0,0,0,0,0)  .
\eeq

Observe that while all expressions in eqs.~(\ref{gamma6t1}) and
(\ref{gamma6t1b}) are known (the explicit form of $\Gamma^{(4)}$ can
be found in paper I, sect.~IIIC), in the r.h.s. of
eq.~(\ref{gamma6t2}) appears the function $\Gamma^{(6)}$, the
variable of the differential equation (\ref{eqcompletegamma62} ). Evaluating all the 15
permutations which includes this  function $\Gamma^{(6)}$, one
verifies that it appears either in the form
$\Gamma^{(6)}(p,-p,0,0,0,0)$, or in the form
$\Gamma^{(6)}(0,0,0,0,0,0)$. The latter being simply the (known) LPA'
expression (see paper I, sect.~IIC), one ends up with a differential
equation for the function $\Gamma^{(6)}(\kappa;p,-p,0,0,0,0)$. In
order to solve it, one needs an initial ansatz for  $\Gamma^{(8)}$
 that  appears in (\ref{gamma6t3}).

\begin{figure}[t!]
\begin{center}
\includegraphics*[width=10 cm,angle=-90]{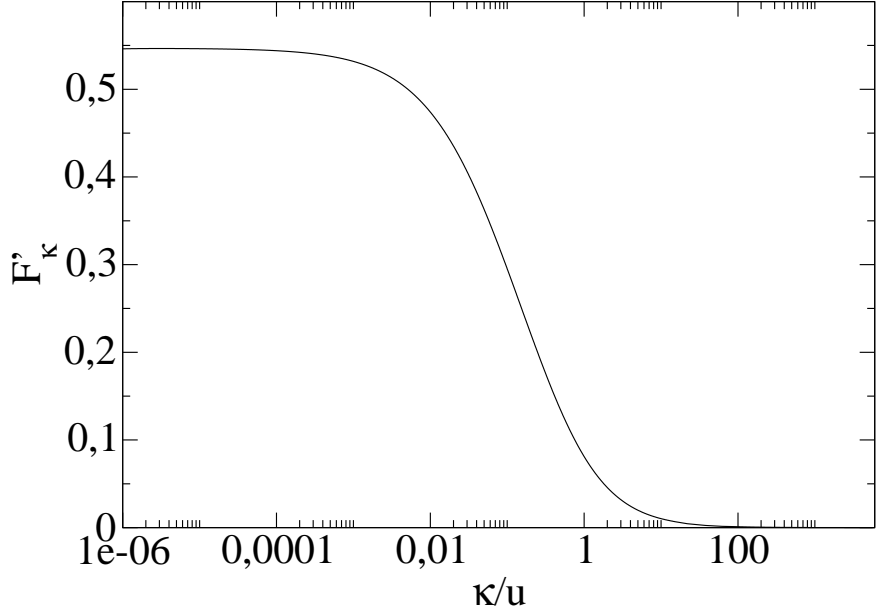}
\end{center}
\caption{\label{Fkp} The function $F_\kappa'$ as a function of
$\kappa/u$.}
\end{figure}

To get the latter, we follow approximation ${\cal A}_3$. Let us write the LPA' equation corresponding
to (\ref{eqcompletegamma62}). This can be obtained by deriving
three times with respect to $\rho$ the equation for the effective
potential, and then setting $\rho=0$. One gets:
\begin{eqnarray}\label{LPA4}
\partial_\kappa V'''(\rho=0)&=&\int \frac{d^dq}{(2\pi)^d} \partial_\kappa R_k(q)G^2(q)\left\{
-3(N+26) G^2(q) \left(V''(\rho=0)\right)^3 \right. \nonumber \\
&+&\left.3(N+14) G(q) V'''(\rho=0)V''(\rho=0)
-\frac{N+6}{2}V^{(4)}(\rho=0) \right\}. \nonumber \\
\end{eqnarray}
Defining (see also paper I, sect. II C)
\beq\label{defL}
V^{(4)}(\rho=0)=l_\kappa=K_d^{-3}Z_\kappa^4 \kappa^{8-3d} \hat
l_\kappa ,
\eeq
 we transform eq.~(\ref{LPA4}) into:
\beq \label{eqpourHk}
 \kappa\partial_\kappa h_\kappa= -3 (N+26) I_d^{(4)}(\kappa) g_\kappa^3
+3(N+14) g_\kappa I_d^{(3)}(\kappa) h_\kappa -\frac{1}{2} (N+6)
 I_d^{(2)}(\kappa)  l_\kappa.
 \eeq
On the other hand, as suggested by  the large $N$ limit
(see paper I, sect. II D), one expects that once approximation
(${\cal A}_2$) is performed, the term containing $\Gamma^{(8)}$ to
be  proportional to the other ones, the coefficient depending only
on $\kappa$. However, while in the case of the equation for
$\Gamma^{(4)}$ the three contributions involving products of
$\Gamma^{(4)}$ in eq.~(\ref{gamma40new}) where identical at zero
external momenta (thus giving a unique proportional factor
$F_\kappa$), here we have two different contributions (those with
three $\Gamma^{(4)}$ and those with one $\Gamma^{(4)}$ and one
$\Gamma^{(6)}$). Thus, the way the contribution of $\Gamma^{(8)}$
can be distributed over the other two is not unique. To remove the
ambiguity, we distribute the various terms as they appear in the
LPA', eq.~(\ref{eqpourHk}): \beq\label{eqFkp} -\frac{1}{2} (N+6)
 I_d^{(2)}(\kappa) l_\kappa=3 (N+26)  I_d^{(4)}(\kappa) F'_\kappa
g_\kappa^3 -3(N+14) I_d^{(3)}(\kappa) F_\kappa g_\kappa h_\kappa,
\eeq where $F_\kappa$ is the function defined in eq.~(I.44), while
eq.~(\ref{eqFkp}) can be taken as the definition of $F'_\kappa$.
Using eqs.~(I.42), (I.44) and (\ref{defL}), one then sets: \beq
F'_\kappa=\frac{(1+\hat m_\kappa^2)^2}{\hat g_\kappa^3(N+26)}
\left[\frac{(N+14)(N+4)}{2(N+8)}\frac{\hat h_\kappa^2}{\hat
g_\kappa} -\frac{N+6}{6}\hat l_\kappa\right].\nonumber\\ \eeq The
function $F'_\kappa$ is shown in fig.~\ref{Fkp}.

We are now in the position to perform the approximation  ${\cal A}_3$ in
the flow equation of $\Gamma^{(6)}$, i.e., in
eq.~(\ref{eqcompletegamma62}). This amounts to the replacement:
\begin{eqnarray}\label{separation}
-\frac{1}{2}\Tr\left\{ \Gamma_{12345678}^{(8)}  G^2\partial_\kappa R_\kappa
\right\}\to \, \, && \hspace{-.9cm} \Tr \left\{ G^2 \partial_\kappa R_\kappa
\left( F'_\kappa \Gamma_{i12j}^{(4)}G
\Gamma_{j34k}^{(4)}G\Gamma_{k56i}^{(4)}+ \mathrm{44\,permutations}
\right.\right.
\nonumber \\
&-& \left. \left. 3(N+14) F_\kappa\Gamma_{i1234j}^{(6)}G
\Gamma_{j56k}^{(4)}+\mathrm{14\,
permutations}\right)\right\}.\nonumber\\
\end{eqnarray}

At this stage, we have all the ingredients to write, and solve,
the (approximate) equation for $\Gamma^{(6)}$. After rewriting
eq.~(\ref{eqcompletegamma62}) with the use of
eq.~(\ref{separation}), and evaluating all the terms as in
eqs.~(\ref{gamma6t1}), (\ref{gamma6t1b}) and (\ref{gamma6t2}), one
ends up  with an ordinary differential equation where the
dependent variable is $\Gamma^{(6)}(\kappa;p,-p,0,0,0,0)$. To
write explicitly this equation it is useful to use the fact that
$\Gamma^{(6)}_{123456}(p,-p,0,0,0,0)$ is completely symmetric
under the exchange of indices 3, 4, 5 and 6. Then, one can make
the decomposition:
\begin{eqnarray}
&&\Gamma_{123456}^{(6)}(p,-p,0,0,0,0)=\Gamma^{(6)}_a
\delta_{12}(\delta_{34}\delta_{56}+\delta_{35}\delta_{46}+\delta_{36}\delta_{45})
\nonumber \\
&+&\Gamma^{(6)}_b \left[
\delta_{24}(\delta_{13}\delta_{56}+\delta_{15}\delta_{36}
+\delta_{16}\delta_{35})
+\delta_{14}(\delta_{23}\delta_{56}+\delta_{25}\delta_{36}+\delta_{26}\delta_{35})
\right. \nonumber \\
&+&\left.
\delta_{13}\delta_{25}\delta_{46}+\delta_{13}\delta_{26}\delta_{45}
+\delta_{15}\delta_{23}\delta_{46}
+\delta_{15}\delta_{26}\delta_{43}+\delta_{16}\delta_{23}\delta_{45}
+\delta_{16}\delta_{25}\delta_{34})
\right]  ,
\end{eqnarray}
Finally, taking the trace over the tensor indices, and doing a
lengthy, but straightforward calculation, one gets:
\begin{eqnarray}
\label{eqgammaa}
 \kappa\partial_\kappa\Gamma^{(6)}_a&=&(1-F_\kappa) I_d^{(3)}(\kappa)
\left\{ g_\kappa
 \left[2(N+8) \Gamma^{(6)}_a + 4 \Gamma^{(6)}_b \right] \right.\nonumber \\
&+&\left. h_\kappa g_{\alpha p}
\left[\left(N+6+\frac{8}{N}\right)\left( \frac{g_\kappa}{g_{\alpha
p}}\right)^{\frac{N+2}{N+8}}
-\frac{8}{N}\left(\frac{g_\kappa}{g_{\alpha p}}
\right)^{\frac{2}{N+8}}\right]\right\} \nonumber \\
&-&3 (1-F'_\kappa) I_d^{(4)}(\kappa) g^2_\kappa g_{\alpha p}
\left\{(N+10+\frac{16}{N})\left(\frac{g_\kappa}{g_{\alpha p}}
\right)^{\frac{N+2}{N+8}}
-\frac{16}{N}\left(\frac{g_\kappa}{g_{\alpha
p}}\right)^{\frac{2}{N+8}}\right\},\nonumber\\
\end{eqnarray}
\beq
\label{eqgammab}
 \kappa\partial_\kappa\Gamma^{(6)}_b&=&(1-F_\kappa) I_d^{(3)}(\kappa) \left\{ g_\kappa (N+16) \Gamma^{(6)}_b
+2 h_\kappa g_{\alpha p}\left(\frac{g_\kappa}{g_{\alpha p}}
\right)^{\frac{2}{N+8}}\right\} \nonumber \\
&-&12 (1-F'_\kappa)I_d^{(4)}(\kappa) g^2_\kappa g_{\alpha p}
\left(\frac{g_\kappa}{g_{\alpha p}}\right)^{\frac{2}{N+8}},\nonumber \\
\eeq
when $\alpha p>\kappa$. When $\alpha p\leq \kappa$, one can show that the LPA' expression
for the 6-point vertex:
\beq
\label{sol1}
\Gamma_{12iijj}^{(6)}=\delta_{12} h_\kappa (N+2)(N+4) \; \;\; (\alpha p\leq \kappa) ,
\eeq
is a solution of the equation that one gets. Then, as it is a first order differential equation in $\kappa$,
the expression (\ref{sol1}) is the solution for the case $\alpha p\leq \kappa$.

Returning to the case $\alpha p>\kappa$, one can diagonalize
eqs.~(\ref{eqgammaa}) and (\ref{eqgammab}), the eigenvalues and
eigenvectors being
\begin{eqnarray}
\lambda_a=2(N+8) &&\hspace{1cm} \mathrm{with\,eigenvector}\, (1,0) \nonumber \\
\lambda_b=N+16 &&\hspace{1cm} \mathrm{with\,eigenvector}\,
(-4/N,1)
\end{eqnarray}
One thus can write:
\beq \label{relgammaeta}
\left(\begin{array}{c}
\Gamma^{(6)}_a\\
\Gamma^{(6)}_b
\end{array}
\right) =a_a\left(\begin{array}{c}
1\\
0
\end{array}
\right) +a_b\left(\begin{array}{c}
-4/N\\
1
\end{array}
\right)
\eeq
where $a_a$ verifies the equation:
\begin{eqnarray}\label{eqaa}
&& \kappa\partial_\kappa a_a=(1-F_\kappa)I_d^{(3)}(\kappa) 2(N+8)g_\kappa a_a \nonumber \\
&&+(1-F_\kappa) I_d^{(3)}(\kappa)
\left(N+6+\frac{8}{N}\right)h_\kappa g_{\alpha p}
\left(\frac{g_\kappa}{g_{\alpha p}}\right)^{\frac{N+2}{N+8}}
\nonumber \\
&&-3(1-{F'}_\kappa)
I_d^{(4)}(\kappa)\left(N+10+\frac{16}{N}\right)\left( g_{\alpha
p}\right)^{3} \left(\frac{g_\kappa}{ g_{\alpha
p}}\right)^{3\frac{N+6}{N+8}}
\end{eqnarray}
while the equation for $a_b$ is simply the equation for
$\Gamma^{(6)}_b$ (\ref{eqgammab}).

Both equations (\ref{eqgammab}) and (\ref{eqaa}) have the following form
\beq
 \kappa\partial_\kappa \Gamma=\lambda I_d^{(3)}(\kappa) (1-F_\kappa)g_\kappa \Gamma+\phi(\kappa)
\eeq its solution being \beq \Gamma(\kappa)=\Gamma(\alpha p)
\left(\frac{g_\kappa}{g_{\alpha p}} \right)^{\frac{\lambda}{N+8}}
+\int_{\alpha p}^\kappa d\kappa' \left(\frac{g_\kappa}{g_{\kappa'}}
\right)^{\frac{\lambda}{N+8}}\phi(\kappa') \eeq

Unfortunately we could not succeed to solve analytically the
equation above, as we did for the initial ansatz for $\Gamma^{(4)}$;
there, the key equation was (I.43), but we could not find a similar
one here.

Imposing the continuity condition, in $\kappa=\alpha p$, dictated by eq.~(\ref{sol1}):
\beq
\Gamma^{(6)}_a(\alpha p)=\Gamma^{(6)}_b(\alpha p)=h_{\alpha p}
\eeq
which gives
\begin{eqnarray}
a_a(\alpha p)&=&\left(1+\frac{4}{N}\right)h_{\alpha p} \nonumber \\
a_b(\alpha p)&=&h_{\alpha p}
\end{eqnarray}
the solutions for $a_a(\kappa)$ and $a_b(\kappa)$ are
\begin{eqnarray}
&&a_a(\kappa)=\left(1+\frac{4}{N}\right)h_{\alpha
p}\left(\frac{g_\kappa}{g_{\alpha p}}
\right)^2 \nonumber \\
&+&\int_{\log \alpha p/\Lambda}^t
dt'\left(\frac{g_\kappa}{g_{\kappa'}}\right)^2
\left\{(1-F_\kappa)I_d^{(3)}(\kappa) (N+6+\frac{8}{N})h_\kappa
g_{\alpha p}
\left(\frac{g_\kappa}{g_{\alpha p}}\right)^{\frac{N+2}{N+8}} \right. \nonumber \\
&-&\left.3(1-F'_\kappa) I_d^{(4)}(\kappa)(N+10+\frac{16}{N})\left(
g_{\alpha p} \right)^{3}
\left(\frac{g_\kappa}{ g_{\alpha p}}\right)^{3\frac{N+6}{N+8}} \right\} \nonumber \\
&&a_b(\kappa)=h_{\alpha p} \left(\frac{g_\kappa}{g_{\alpha p}}
\right)^{\frac{N+16}{N+8}} \nonumber \\
&+&\int_{\log \alpha p/\Lambda}^t
dt'\left(\frac{g_\kappa}{g_\kappa'} \right)^{\frac{N+16}{N+8}}
\left\{2 H_\kappa g_{\alpha p}\left(\frac{g_\kappa}{g_{\alpha p}}
\right)^{\frac{2}{N+8}}\right. \nonumber \\
&-&\left. 12 (1-F'_\kappa)I_d^{(4)}(\kappa) g^2_\kappa g_{\alpha p}
\left(\frac{g_\kappa}{g_{\alpha p}}\right)^{\frac{2}{N+8}}\right\} \nonumber \\
\end{eqnarray}

Finally, when $\kappa < \alpha p$, one has
\beq\label{sol2}
\Gamma_{12iijj}^{(6)}(p,-p,0,0,0,0)=\delta_{12}N(N+2)a_a(\kappa) \;\; (\kappa < \alpha p) .
\eeq

\section{Products of functions $\tilde\Gamma^{(4)}$ in eq.~(\ref{4pointLOapp})}

\subsection{The $s$ and $t$-channel contributions}

\label{sun-tad}

In this  appendix we obtain  explicit expressions for the products of functions $\tilde\Gamma^{(4)}$ that appear in the r.h.s. of eq.~(\ref{4pointLOapp}), where $\tilde\Gamma^{(4)}$ is the  initial ansatz for the  4-point function.  All the needed expressions for the 4-point fucntions that are needed here are those for  $\tilde\Gamma^{(4)}(p_1,p_2,0,-p_1-p_2)$ that can be found in paper I, sect. III B.

{\bf 1) The $s$-channel contribution}

Here we consider the product
$\tilde\Gamma^{(4)}_{12ij}(\kappa; p,-p,0,0)\tilde\Gamma^{(4)}_{llij}(\kappa; q,-q,0,0)$, where,  because of the
regulator, $q<\kappa$. There are  two regions to examine:

{\bf a) $\alpha p <\kappa $}

Both 4-point functions are in the region (a) of paper I, sect. III C.
After a simple calculation one gets:
\beq
\tilde\Gamma^{(4)}_{12ij}(p,-p,0,0)\tilde\Gamma^{(4)}_{llij}(q,-q,0,0)=(N+2)^2 g_\kappa^2 \delta_{12} .
\eeq

{\bf b) $\alpha p > \kappa $}

Here, while $\tilde\Gamma^{(4)}_{llij}(q,-q,0,0)$ is still in region (a), the other vertex
is in region (b) of paper I, sect. III C. In this case one gets:
\beq
\tilde\Gamma^{(4)}_{12ij}(p,-p,0,0)\tilde\Gamma^{(4)}_{llij}(q,-q,0,0)=(N+2)^2
g_\kappa g_{\alpha p}  \left(\frac{g_\kappa}{g_{\alpha
p}} \right)^{\frac{N+2}{N+8}} \delta_{12} .
\eeq

{\bf 2) The $t$-channel contribution}

Here we consider the product
$\tilde\Gamma^{(4)}_{12ij}(p,-p,0,0)\tilde\Gamma^{(4)}_{llij}(q,-q,0,0)$. As $q<\kappa$ and
$\alpha <1$ (and thus $\alpha q \le \kappa$), one has  two cases to study:
$p >|p+q|$ and $|p+q|>p$.

{\bf A) $p > |p+q|$ }

In paper I, sect. III B,  we assumed $p_1>p_2>|p_1+p_2|$. Using the
symmetry of the bosonic $n$-point functions, we can conveniently
rewrite the product as
\begin{eqnarray}
&&\tilde\Gamma^{(4)}_{1lij}(p,q,0,-p-q)\tilde\Gamma^{(4)}_{2lij}(-p,-q,0,p+q)=
\tilde\Gamma^{(4)}_{1jil}(p,-p-q,0,q)\tilde\Gamma^{(4)}_{2jil}(p,-p-q,0,q) , \nonumber \\
\end{eqnarray}
and the expressions of the vertices to consider are those of either regions
(a), (b), or (c) of paper I, sect. III B, depending on the value of
$\kappa$ (region (d) never enters, because $q <\kappa$):

{\bf a) $\kappa > \alpha p$}

The two vertices are in region (a) of paper I, sect. III B. The product is simply:
\beq\label{appgaa}
\tilde\Gamma^{(4)}_{1jil}(p,-p-q,0,q)\tilde\Gamma^{(4)}_{2jil}(p,-p-q,0,q)=3(N+2)
g^2_\kappa \delta_{12} .
\eeq

{\bf b) $\alpha p> \kappa > \alpha |p+q|$}

Now, both vertices are in region (b) of paper I, sect. III B. A
lengthy but straigtforward calulation yields:
\begin{eqnarray}\label{appgab}
&&\tilde\Gamma^{(4)}_{1jil}(p,-p-q,0,q)\tilde\Gamma^{(4)}_{2jil}(p,-p-q,0,q)=\nonumber \\
&&g^2_{\alpha p}\frac{N+2}{N^2+4N+20}\left\{
\left(\frac{3}{2} N^2 +6N +30 + \frac{N+14}{2} \sqrt{N^2 +4N+20}
\right)
\left(\frac{g_\kappa}{g_{\alpha p}}\right)^{\frac{2\lambda_+}{N+8}}
\right. \nonumber \\
&&\left.+\left(\frac{3}{2} N^2 +6N +30 - \frac{N+14}{2} \sqrt{N^2
+4N+20} \right)
\left(\frac{g_\kappa}{g_{\alpha p}}\right)^{\frac{2\lambda_-}{N+8}} \right\}
\delta_{12} \nonumber \\
\end{eqnarray}

{\bf c) $\alpha |p+q|>\kappa$}

Both vertices are now in region (c) of paper I, sect. III B. After another
straigtforward calulation one obtains:
\begin{eqnarray}\label{appgac}
&&\tilde\Gamma^{(4)}_{1jil}(p,-p-q,0,q)\tilde\Gamma^{(4)}_{2jil}(p,-p-q,0,q)=  \nonumber \\
&&\left\{ N\left(b^+_{\alpha |p+q|}
\left(\frac{g_\kappa}{g_{\alpha
|p+q|}}\right)^{\frac{\lambda_1}{N+8}} \right.
\nonumber +b^-_{\alpha |p+q|}\left(\frac{g_\kappa}
{g_{\alpha |p+q|}}\right)^{\frac{\lambda_2}{N+8}}\right)^2  \nonumber \\
&&-2N b^-_{\alpha |p+q|}\left(\frac{g_\kappa}{g_{\alpha |p+q|}}
\right)^{\frac{\lambda_2}{N+8}} \nonumber \\
&&\left(b^+_{\alpha |p+q|} \left(\frac{g_\kappa}{g_{\alpha
|p+q|}}\right)^{\frac{\lambda_1}{N+8}}
+b^-_{\alpha |p+q|}\left(\frac{g_\kappa}{g_{\alpha |p+q|}}\right)^
{\frac{\lambda_2}{N+8}}\right) \nonumber \\
&&\left.+\frac{N^2}{2}(N+1)b^-_{^2(\alpha
|p+q|}\left(\frac{g_\kappa}{g_{\alpha
|p+q|}}\right)^{2\frac{\lambda_2}{N+8}}
 +2 (N-1) {\Gamma^C}^2_{\alpha |p+q|}
\right\} \delta_{12} \nonumber \\
\end{eqnarray}
where $b^+_{\alpha |p+q|}$, $b^-_{\alpha |p+q|}$ and
$\Gamma^C_{\alpha |p+q|}$ follow from eqs.~(I.92) and (I.93)
respectively.

{\bf B) $p \le |p+q|$ }

In this case, we reorder momenta and indices as:
\beq
\tilde\Gamma^{(4)}_{1lij}(p,q,0,-p-q)\tilde\Gamma^{(4)}_{2lij}(-p,-q,0,p+q)=
\tilde\Gamma^{(4)}_{j1il}(-p-q,p,0,q)\tilde\Gamma^{(4)}_{j2il}(-p-q,p,0,q) .
\eeq

Similarly as in previous case (A) one has three regions, $\kappa > \alpha |p+q|$,
$\alpha |p+q|> \kappa > \alpha p $ and $\alpha p>\kappa$. For each region the result
is the same as those in eqs.~(\ref{appgaa}), (\ref{appgab}) and (\ref{appgac}), but
exchanging $p$ with $|p+q|$.